\documentclass[12pt,preprint]{aastex}

\pdfoutput=1

\usepackage{epsfig}
\usepackage{natbib}
\def\plotone#1{\centering \leavevmode
\includegraphics[clip=, width=.85\columnwidth]{#1}}
\def\plottwo#1#2{\centering \leavevmode
\includegraphics[width=.45\columnwidth]{#1} \hfil
\includegraphics[width=.45\columnwidth]{#2}}
\def\plotthree#1#2#3{\centering \leavevmode
\includegraphics[width=.3\columnwidth]{#1} \hfil
\includegraphics[width=.3\columnwidth]{#2} \hfil
\includegraphics[width=.3\columnwidth]{#3}}

\newcommand{\cN}[1]{\mathcal{N}}

\def\gsim{\;\rlap{\lower 2.5pt
 \hbox{$\sim$}}\raise 1.5pt\hbox{$>$}\;}
\def\lsim{\;\rlap{\lower 2.5pt
   \hbox{$\sim$}}\raise 1.5pt\hbox{$<$}\;}

\begin{document}


\title{
Habitable Climates}

\author{David S. Spiegel\altaffilmark{1}, Kristen Menou\altaffilmark{1},
Caleb A. Scharf\altaffilmark{1,2}}

\affil{$^1$Department of Astronomy, Columbia University, 550 West 120th Street, New York, NY 10027}
\affil{$^2$Columbia Astrobiology Center, Columbia University, 550 West 120th Street, New York, NY 10027}

\vspace{0.5\baselineskip}

\email{dave@astro.columbia.edu, kristen@astro.columbia.edu,
  caleb@astro.columbia.edu}

\begin{abstract}
According to the standard liquid-water definition, the Earth is only
partially habitable.  We reconsider planetary habitability in the
framework of energy-balance models, the simplest seasonal models in
physical climatology, to assess the spatial and temporal habitability
of Earth-like planets. We quantify the degree of climatic habitability
of our models with several metrics of fractional habitability.
Previous evaluations of habitable zones may have omitted important
climatic conditions by focusing on close Solar System analogies. For
example, we find that model pseudo--Earths with different rotation
rates or different land--ocean fractions have fractional
habitabilities that differ significantly from that of the Earth
itself.  Furthermore, the stability of a planet's climate against
albedo-feedback snowball events strongly impacts its
habitability. Therefore, issues of climate dynamics may be central in
assessing the habitability of discovered terrestrial exoplanets,
especially if astronomical forcing conditions are different from the
moderate Solar System cases.
\end{abstract}


\section{Introduction}
\label{sec:intro}

Planetary science is being challenged by extrasolar planetary system
discoveries \citep{mayor+queloz1995,marcy+butler1996}.\footnote{See
\citet{reipurth_et_al2007} for a review.  See also http://exoplanet.eu
and http://exoplanets.org.} The orbital architectures of many of these
systems, with massive planets located either very near to their parent
stars or on highly eccentric orbits, are strikingly unusual. These
discoveries have been surprising, however, mostly as a consequence of
our own Solar-System centric point of view. In fact, planet searches
at multi-AU orbital distances are now providing tentative evidence
that Solar System giant planets, with their nearly circular orbits,
are more the exception than the norm, by simple comparison with the
ensemble properties of known extrasolar planetary systems
\citep{marcy_et_al2005}.  If these preliminary trends hold, they may
have a profound impact on our perception of our place in the Universe.

The pace of exoplanet discoveries has accelerated sharply in the last
few years and the future looks bright. Two significant developments in
exoplanet research have occurred within the last year or so. A
micro-lensing discovery of a likely terrestrial planet, with a mass
$5.5$ times that of Earth, was reported by \citet{beaulieu_et_al2006},
making it the first discovery of a planet that is thought to be
terrestrial (i.e. rocky). Several months ago, \citet{udry_et_al2007}
announced the Doppler-velocimetry discovery of a potentially habitable
terrestrial planet around a low-mass M-dwarf.  In considering future
discoveries, of specific interest to this study are two dedicated
space missions: {\it Corot} and {\it Kepler}
\citep{baglin2003,borucki_et_al2003,borucki_et_al2007}. A key
objective of both missions is to monitor a large number of stars to
detect the (repeatable) micro-eclipses generated when terrestrial
planets transit in front of their host star.  It is expected that a
few Earth analogues (i.e. with Earth-like masses and comparable
distances to their host stars), and possibly hundreds to thousands of
additional terrestrial planets unlike Earth, will be identified by
{\it Corot} and {\it Kepler} after a few years of operation
\citep{borucki_et_al2007, borucki_et_al2003, basri_et_al2005}. The
discovery of an Earth-like exoplanet, potentially hosting life as we
know it, is therefore within the 5-year astronomical
horizon. Ambitious missions are also being prepared to map in detail
the orbits of exoplanets around nearby stars (with {\it SIM
PlanetQuest} \citep{unwin_et_al2007}) and later to obtain spectra of
nearby Earth-like planets in the hope that they would reveal the first
unambiguous signatures of life on a remote world (with the {\it
Terrestrial Planet Finder} and {\it Darwin} \citep{leger+herbst2007}).

The idea that liquid surface water is a prerequisite for a terrestrial
planet to have the ability to host life is widely used as the key
concept behind searches for habitable planets around other stars.
This is because of the central and critical role that water plays in
the biochemistry on Earth.\footnote{It should however be noted that
alternate molecules (e.g., ammonia, methyl alcohol) could conceivably
perform equivalent roles in different environments, such as those of
lower temperature and higher pressure
\citep{haldene1954,firsoff1963,goldsmith+owen2002}.}  As the current
exoplanet census indicates, notions based on an old Solar System
centric view may only be relevant to a minority of planets (or
planetary systems). What if most terrestrial planets discovered in the
future have, like the vast majority of exoplanets currently known,
highly eccentric orbits generating large seasonal variations?  What if
the atmospheric mass and composition, planetary spin rate, or land-sea
coverage of these exo-Earths are generally different from what they
are on Earth?  If the last ten years of extrasolar planet discoveries
offer any guidance, our Solar System appears to show but little of the
general planetary diversity found around other stars.

The classical calculations of habitability on Earth-like planets by,
e.g., \citet{dole1964} and \citet{hart1979} predated extrasolar
discoveries. Apart from a few exceptions
(\citealt{franck_et_al2000b,gaidos_et_al2005}; see also
\citealt{vonbloh_et_al2007,selsis_et_al2007} in the specific context
of the Gleise 581 system) the subject of planetary atmospheric
habitability has been revisited little since the seminal work of
\citet{kasting_et_al1993}. Given the major developments expected in
the next five years and beyond, it is important that the climate
regimes expected on exotic versions of the Earth, and their
consequences for habitability, be studied and better understood.  This
will help interpret upcoming planet detections and will inform future
long-term efforts on the best strategies to find robust signatures of
life on exo-Earths.

The central role of astronomical forcings in determining the seasons
and the climatology of the Earth is well known (e.g., the long-term
Milankovitch cycles).  Clearly, obliquity, precession, and
eccentricity can all strongly affect global and regional habitability
conditions on the Earth and, by extension, on any other potentially
habitable terrestrial exoplanet.  Previously, only preliminary
investigations of the role of obliquity
(\citealt{williams+kasting1997}, hereafter WK97; and
\citealt{williams+pollard2003}) and eccentricity
\citep{williams+pollard2002}, for very specific Earth-like conditions,
have been considered in some detail. The surface habitability of a
terrestrial planet, however, must depend on the combination of
obliquity and eccentricity with the planetary rotation rate, the
continental coverage (from dry Earths to water worlds) and the overall
mass and composition of the atmospheric layer (e.g., compare Mars
vs. Venus), among other factors.  A thorough exploration of how these
various global planetary attributes combine to affect climatology and
make a terrestrial planet seasonally or regionally habitable is
therefore an important element in the search for signatures of life
elsewhere. In the present study, we describe how energy balance
climate models can contribute to this understanding.

In \S~\ref{sec:model}, we introduce the concept of climate modeling
hierarchy, describe our energy balance model and validate it on the
Earth.  In \S~\ref{sec:Hab Zones}, we reconsider various features of
habitability for Earth--like planets with seasonally-forced climates.
In \S~\ref{sec:def}, we discuss several subtleties in the definition
of habitability that emerge from our work on seasonal climates, and we
finally conclude in \S~\ref{sec:conc}.

\section{Climate Modeling Hierarchy}
\label{sec:model}

Current modeling tools to study the Earth's climate are constructed to
be specific to Earth.  For example, their radiative transfer schemes
are elaborate and specific to the conditions of Earth's atmosphere and
are valid only within fairly narrow ranges of atmospheric composition,
temperature and pressure. Their surface boundary conditions are also
specific to a given land-ocean-ice configuration.  The same is
essentially true of advanced climate tools used to model the
atmospheres of Mars or Venus. This strongly limits the region of
planetary parameter space that can be explored with these
computationally intensive tools.  Such models are bound to produce a
somewhat limited view of what constitutes a habitable planet.  They
have been useful in providing some initial insights on astronomical
factors affecting habitability (e.g.,
\citealt{williams+pollard2002,williams+pollard2003}), but given the
likely detection of terrestrial exoplanets with various global
attributes in the near future, it is important that tools addressing
the specific needs of astronomers be developed now, to help assess the
potential habitability of discovered exoplanets.

A difficult aspect of assessing the habitability of discovered
terrestrial exoplanets will be that many of the key planetary
attributes that determine the climate regime will remain entirely
unknown to astronomers, at least in the foreseeable future. Consider,
for instance, the terrestrial planets that will be discovered by {\it
Kepler}. {While their semimajor axes, their radii,} and thus
indirectly their masses and surface gravities (e.g.,
\citealt{valencia_et_al2006,valencia_et_al2007,fortney_et_al2007,adams_et_al2007})
may be reasonably well constrained by the data -- but see
\citet{seager_et_al2007} -- very little additional information on the
global planetary attributes will be available to help us evaluate
their potential habitability. In some special cases, it may be
possible to set constraints on orbital eccentricities
\citep{barnes2007,ford_et_al2008}, but in general, except for
tidally-locked planets, the eccentricity, obliquity and planetary
rotation rate will all be essentially unknown. Perhaps even more
important, the atmospheric mass and composition of detected planets
will be unknown. Even though an incremental amount of information on
rocky exoplanets will become gradually available with future
generations of space missions (i.e., orbits with {\it SIM-PlanetQuest}
and later spectra with {\it TPF} and {\it Darwin}), this bright future
does not alleviate the need for an interpretation of {\it Kepler},
COROT, and radial velocity exoplanet discoveries in terms of
habitability.

To address this interpretational challenge, one approach is to
consider the climate problem at a fundamental level, using first
principles to systematically classify planets whose climate regimes
and associated surface physical conditions might permit
habitability. As we shall see below, an approach based on
1-dimensional Energy Balance Models (EBMs) is particularly attractive
because it is computationally efficient.  This approach, therefore,
permits a relatively thorough exploration of the multi--dimensional
parameter space of planetary properties in order to identify the most
promising regions of that space for habitability.

In physical climatology (e.g., \citealt{hartmann1994,ghil2002}), a
hierarchical family of climate models has been built from tools of
increasing complexity. At the lowest level of the hierarchy,
quantitative descriptions of planetary climate start with simple,
global radiative balance models.  These models focus on a steady
radiative equilibrium solution, associate a single surface temperature
to an entire planet, and ignore important equator-to-pole dynamical
atmospheric fluxes. They cannot account for the time--variable
property, nor the regional property of climate and habitability. It is
significant that, even though much of Earth itself is only transiently
or regionally habitable by the standard liquid water criterion,
calculations of habitability for astronomical applications have almost
exclusively relied on global radiative balance models, of the type
described in \citet{kasting_et_al1993}.\footnote{Note that global
radiative models do account for 1 dimension, in the vertical, by
describing the vertical radiative-convective structure of the studied
atmosphere, in an average sense.}  As a result, existing work on
habitability has largely emphasized radiative and chemical issues.
Tackling the equally important dynamical ones requires regional and
seasonal climate models.

At the next level of the hierarchy, 1-dimensional EBMs solve a 1D
time-dependent diffusion equation to specify the evolution of the
surface temperature as a function of planetary latitude, based on
seasonal variations in incoming and outgoing radiative fluxes and the
energetic redistribution due to atmospheric motions. In the presence
of a thick enough atmosphere, dynamical transport is indeed an
important contributor to the local thermal budget of a terrestrial
planet\footnote{Atmospheric motions result precisely from an
atmosphere being locally out of radiative equilibrium, even when
radiative equilibrium is globally satisfied.} and therefore affects
its regional habitability properties. Ever since the seminal work of
\citet{budyko1969} and \citet{sellers1969} on the Earth's climate, 1D
EBMs have been recognized as useful tools in physical climatology,
especially for studies of variations in external (astronomical)
forcing conditions and their effects on climate stability
\citep{hartmann1994}. Except for a few preliminary investigations
(e.g., WK97; \citet{franck_et_al2000b,gaidos+williams2004}), however,
1D EBMs have generally not been applied to the outstanding problem of
habitability on terrestrial exoplanets.

In the present work, we describe EBMs for planets that are largely
similar to Earth, partly because this permits us to validate our new
modeling tool on this well known and understood case.  In the course
of presenting our models, we will use two terms, ``Earth--like'' and
``pseudo--Earth'', to distinguish between model planets that are,
within the context of our 1-dimensional energy balance framework, as
similar to Earth as we can achieve with very simple heating and
cooling functions (``Earth--like'') and model planets that are less
specifically tuned to Earth by virtue of, e.g., having a different
rotation rate (``pseudo--Earth'').  We investigate how the
habitability of our pseudo--Earths varies with their distance from a
sun--like star, and with the efficiency of the latitudinal
redistribution of heat. In this study, we consider only cases with 0
orbital eccentricity and an Earth-like 23.5-degree obliquity.  As
illustrated below, we already find that the issue of habitability for
close Earth analogs becomes a rich one when addressed with a 1D
time-dependent EBM.

\subsection{Global Radiative Balance}
\label{ssec:globrad}

It is instructive to start our investigation of habitable climates
with a discussion of global radiative balance results. Assuming that
the annual mean surface temperature, $T$, on an Earth--like planet is
determined by radiative balance between energy sources and sinks, the
following steady-state radiative equilibrium equation must be
satisfied,
\begin{equation}
\label{eq:rad eq}
I[T] = S (1 - A[T]),
\end{equation}
where $I$ is the infrared cooling flux, $S$ is the annual mean stellar
insolation flux and $A$ is the global planetary albedo.

In this work, we consider three different formulations for the cooling
and albedo functions. Compared to previous work on global radiative
balance \citep[e.g.][]{kasting_et_al1993}, our treatment is extremely
simple. Nevertheless, it accounts for the key physical elements
involved in the absorption and reemission of incident stellar flux and
is sufficient for our EBM work, which is focused on dynamical climate
issues rather than radiative transfer ones.

In a planetary atmosphere, the greenhouse effect acts to reduce the
infrared cooling flux, $I[T]$, at a given value of the surface
temperature, $T$.  \citet{kasting1988} and \citet{kasting_et_al1993},
for instance, describe in detail how the Earth's greenhouse effect
scales with surface temperature.  An important result is that higher
temperatures cause greater humidity, which in turn leads to stronger
atmospheric heat retention via the greenhouse effect from water vapor
absorption (see also WK97). Our purpose here is not to accurately
reproduce Earth's climate but rather to investigate dynamic climate
issues for Earth-like planets. We therefore consider three different
models for the infrared cooling function, $I$, and the albedo
function, $A$, as listed in Table~\ref{tab:one}.  While each of these
models reproduces reasonably well the current climate of the Earth,
(see \S~\ref{sssec:valid}), they could each respond differently once we
start exploring physical conditions that are different from the
current Earth. These three atmosphere models are therefore useful in
evaluating the sensitivity of our main results on habitable climates
to detailed model assumptions for the infrared cooling and albedo
functions.

Models~1 and~2 have infrared cooling functions inspired from a
one-zone radiative transfer formulation that assumes ground-level
blackbody emission ($\sigma T^4$), effectively reduced at the
atmospheric ``photosphere'' according to a near-Eddington
approximation (inspired by \citet{shu1982}), for a given atmospheric
optical thickness to infrared radiation, $\tau_{\rm IR}$ (as shown in
Table~\ref{tab:one}). Model~3, on the other hand, adopts the standard
linear cooling function of \citet{north+coakley1979}, which can be
interpreted as a linearization of IR--cooling around the
globally-averaged conditions for the Earth.  It is in close agreement
with other T-linearized IR models used in the geophysical literature
\citep[e.g.,][]{budyko1969,north_et_al1981}. In our model~1, the
IR optical thickness is fixed to unity, while in model~2, a cubic
dependence on surface temperature, $T$, is adopted, to better match
the linear model results (model~3) for Earth-equivalent conditions.

In each of our three atmospheric models, the albedo function, $A[T]$,
is chosen to capture a rapid ice-water transition, with values of $A
\sim $\,0.7--0.77 well below 263~K (ice--covered surface), and values
$\sim $\,0.25--0.3 well above 273~K (ice--free surface). To avoid
albedo discontinuities, which are known to result in spurious ``small
ice-cap'' instabilities in EBMs (e.g., \citet{held_et_al1981}), the
albedo transition is performed smoothly over the range 263--273~K with
a hyperbolic tangent formulation. This type of albedo prescription is
standard in the context of EBMs (e.g. WK97). The slightly different
asymptotic values adopted for the albedos in our three models (see
Table~\ref{tab:one}) were adjusted to best reproduce the annual mean
climate of the Earth. This is common practice with simplified EBMs
which do not account for the detailed land-surface conditions that
contribute to the global planetary albedo (see \S~\ref{sssec:limits}
for a discussion of EBM limitations).

In what follows, we adopt model~2 as our fiducial model but we also
show a few comparison results with models~1 and~3. Quite generally, we
find that predictions from our three atmospheric models differ only
modestly at a quantitative level. With a choice of IR-cooling and
albedo functions, steady-state radiative equilibrium solutions for the
annual mean climate are obtained by solving Eq.~(\ref{eq:rad eq}) for
a specified value of the annual mean insolation flux, $S$, received by
the planet. For a fast-spinning planet, such that a diurnal average is
justified, on a circular orbit at a distance $a$ (in AU) from a
Sun-like star, the annual mean insolation is simply $S=(1/4) \,
S_0/a^2$ (e.g. \citealt{rubincam2004}), where the solar constant
$S_0=1.36 \times 10^6$~erg~cm$^{-2}$~s$^{-1}$.

\begin{figure}[h!]
\plotone{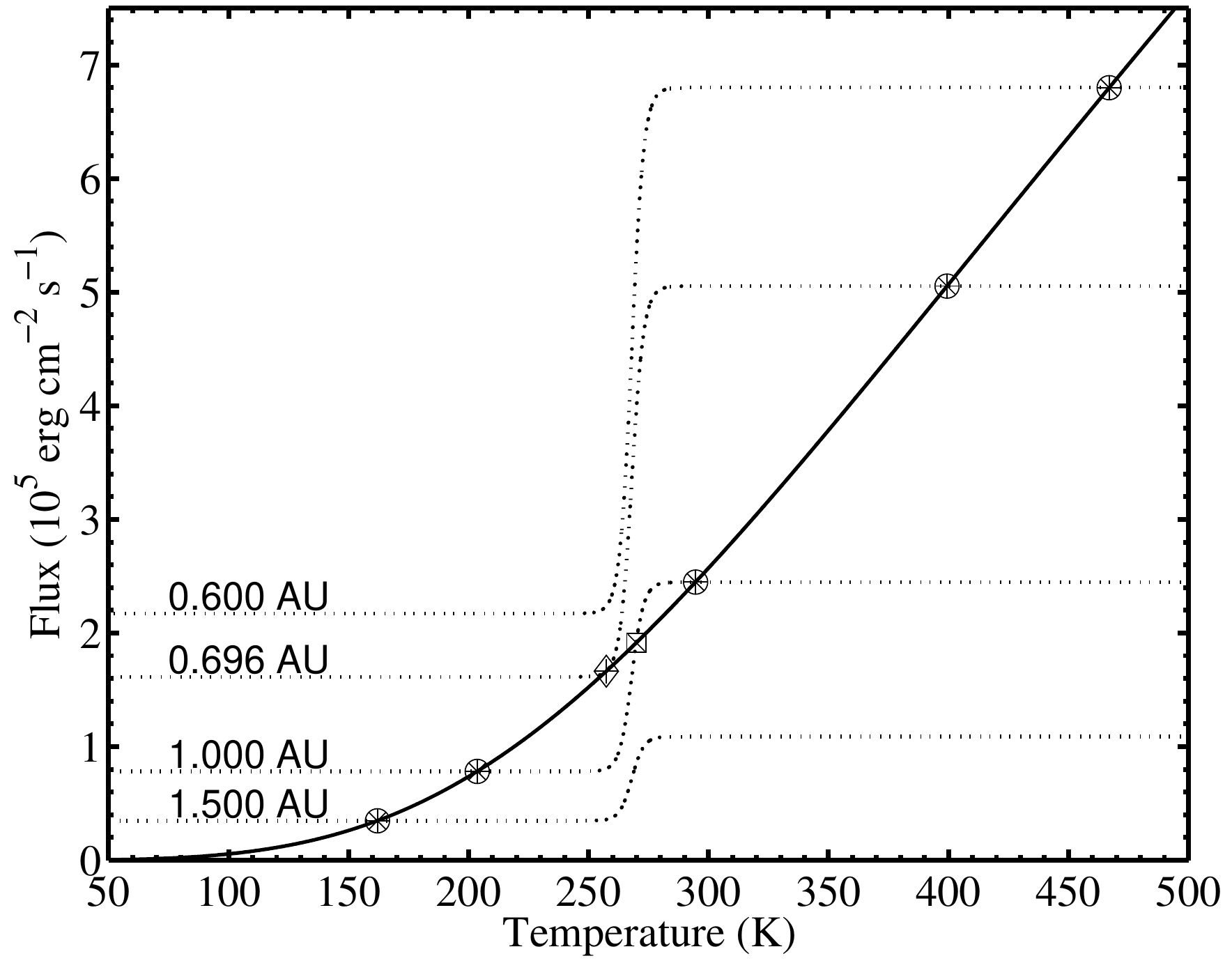}
\caption{Heating and cooling fluxes, as a function of surface
temperature, in a global radiative balance model using the IR-cooling
and albedo functions listed as model~2 in Table~\ref{tab:one}.  The
solid line shows the IR-cooling flux while dotted lines show
albedo-reduced, annual mean insolation fluxes at four different
orbital distances from a Sun--like star (from 0.6~AU to 1.5~AU).
Stable radiative equilibrium solutions are highlighted with
circles. Other solutions are thermally unstable (see text for
details). In the vicinity of $\sim 1$~AU, this class of global
radiative balance models generally admits two stable solutions for the
climate; one is ice-free and the other is ice-covered.}
\label{fig:mult sols}
\end{figure}

With these assumptions, Eq.~(\ref{eq:rad eq}) reduces to the thermal
balance equation $Q^-=Q^+$, where the annual mean cooling rate,
$Q^-=I[T]$, and heating rate, $Q^+ = (1-A[T])S$, are expressed per
unit surface area of the planet. Due to the steep albedo variation
with surface temperature, relative to the smooth IR-cooling
dependence, it is possible for multiple climate equilibrium solutions
to satisfy Eq.~(\ref{eq:rad eq}).  Figure~\ref{fig:mult sols}
illustrates this possibility, using our model~2 for definiteness. The
cooling and heating rates, $Q^-$ and $Q^+$, are shown as solid and
dotted lines, respectively, for a wide range of surface
temperatures. The heating rate is shown at four different orbital
distances from the central Sun-like star, from $0.6$ to $1.5$~AU. The
strong effect of the albedo transition on the heating rate is easily
identified at $T \simeq 263$--$273$~K.

While unique climate equilibrium solutions satisfy Eq.~(\ref{eq:rad
eq}) for $a = 0.6$ and $1.5$~AU, two such solutions are found at $a =
0.696$~AU and three solutions are found at $a = 1$~AU, as
shown by the intersections of dotted and solid lines in
Fig.~\ref{fig:mult sols}. Solutions to a thermal balance equation such
as Eq.~(\ref{eq:rad eq}) are stable only if {small increases in
temperature cause cooling to exceed heating and small decreases cause
heating to exceed cooling:}
\begin{equation}
\frac{d \left( Q^+ -Q^- \right)}{dT} <0,
\end{equation}
If the above equation is satisfied, then any small surface temperature
perturbation relaxes back to the starting equilibrium
solution. Figure~\ref{fig:mult sols}, then, shows that solutions found
in the steep part of the albedo transition ($T \simeq 263$--$273$~K)
are generally unstable, given the relatively slow increase of the
IR-cooling flux with surface temperature.  The intermediate
temperature solution in the case $a = 1$~AU is therefore unstable and
the lower temperature solution in the marginal case $a = 0.696$~AU is
stable only on one side (stable against reductions in temperature but
not against increases).

At small or large enough orbital distances, a unique stable solution
exists, corresponding to either an ice-free or an ice-covered
climate. At intermediate orbital distances ($\sim 1$~AU), however,
both the ice-free and the ice-covered climate solutions are valid and
stable. This bi-stable property of Earth's climate in simple global
radiative balance models is well-known and it may be related to the
snowball (globally ice-covered) events for which there is some
evidence in the past history of Earth's climate (see the review by
\citet{hoffman+schrag2002} and references therein). We also note that
these ice-covered equilibrium solutions correspond to the
``cold-start'' scenarios briefly mentioned by \citet{kasting_et_al1993}
in their detailed study of habitability with global radiative balance
models.

Globally ice-covered climate solutions are often omitted from studies
of planetary habitability because they violate the standard surface
liquid water requirement for habitability. Yet, Earth's climate itself
may have experienced one or two such global glaciation events and
recovered from them without catastrophic damage to life on the planet,
or at least without eradicating life from the planet
\citep{hoffman_et_al1998,hoffman+schrag2000,baum+crowley2003,hoffman+schrag2002}.
The fact that the current Earth is partially ice--covered, depending
on the seasons, reveals another obvious shortcoming of global
radiative balance calculations for habitability: an Earth-like planet
can be habitable even if it is regionally and/or transiently
ice-covered. This important class of climate regime is simply not
addressed by global radiative balance models admitting only fully
ice-covered or fully ice-free stable solutions, as illustrated in
Fig.~\ref{fig:mult sols}. The present work on seasonally habitable
climates is motivated precisely by the possibility to explore these
more complex climate regimes with a simple class of climate models.

\subsection{1--D Energy Balance Model}
\label{ssec:EBM}

Issues of seasonal climate dynamics and susceptibility to climate
transitions are not addressed by steady--state radiative balance
models that consider only annually averaged global mean conditions.
In principle, one can solve for the instantaneous radiative
equilibrium conditions either locally or globally on an Earth--like
planet.  A planet satisfying radiative equilibrium globally, however,
does not necessarily satisfy it locally, for two main reasons: First,
atmospheric motions tend to transport heat from the hotter regions to
the colder regions, effectively coupling the thermal states of various
locations on the planet beyond a simple state of radiative balance.
Second, the atmosphere has thermal inertia and does not immediately
adjust its temperature to the local forcing.  As a result, local
radiative equilibrium solutions will generally not be accurate
representations of local surface temperature conditions.  Furthermore,
because of this thermal inertia, planets do not necessarily satisfy
even global radiative equilibrium, especially when on eccentric
orbits.  These important physical processes can be incorporated in a
time--dependent EBM that models the seasonal climate as a latitudinal
diffusion of atmospheric heat under prescribed astronomical forcing.

A justification of the equivalent diffusion equation satisfied by an
atmospheric layer, obtained by averaging the second law of
thermodynamics and the continuity equation, can be found in
\citet{lorenz1979}. \citet{lorenz1979} also argues that the diffusion
approximation is justified for the Earth, on large enough scales for
the atmosphere to be considered a forced system. Following WK97 and
previous work in the geophysical literature, we adopt the following
prognostic diffusion equation for the planetary surface temperature,
\begin{equation}
\label{eq:diffu eq1}
C \frac{\partial T[x,t]}{\partial t} - \frac{\partial}{\partial x}
\left( D(1 - x^2) \frac{\partial T[x,t]}{\partial x} \right) + I[T] =
S (1 - A[T]),
\end{equation}
where all quantities are local functions, $x \equiv \sin \lambda$ is
the sine of the latitude $\lambda$, $T$ is the surface temperature,
$C$ is the effective heat capacity of the atmosphere (different over
land, ocean, and ice), $D$ is the diffusion coefficient that
determines the efficiency of latitudinal redistribution of heat, $I$
is the same IR-cooling flux as in \S\ref{ssec:globrad}, $S$ is the
diurnally-averaged insolation flux and $A$ is the same albedo function
as in \S\ref{ssec:globrad}. In the above equation, $C$, $D$, $I$, and
$A$ are functions of $T$, $x$, $t$, and possibly other relevant
planetary parameters. The $(1 - x^2)$ factor multiplying the diffusion
coefficient in Eq.~(\ref{eq:diffu eq1}) is a standard prescription to
capture the reduced efficiency of latitudinal heat transport in the
polar regions of an Earth-like planet (see, e.g.,
\citet{north_et_al1983}).

Our prescriptions for $C$ and $D$ are borrowed from WK97 and the
existing geophysical literature on 1D EBMs. For simplicity and
flexibility, we use simple, physically motivated prescriptions.  For
the surface heat capacity $C$, we assume a uniform ocean fraction of
$70\%$ at every latitude and use the same partial heat capacities over
land, ocean, and ice as WK97: $C_l = 5.25 \times 10^9\,{\rm
erg~cm^{-2}~K^{-1}}$ over land, $C_o = 40\, C_l$ above the wind-mixed
surface layer of the ocean, $C_i = 9.2 \, C_l$ over ice when $263 < T
< 273 {\rm ~K}$, and $C_i = 2.0 \, C_l$ for $T \leq 263 {\rm~K}$ (see
WK97 for details).  For the diffusion coefficient $D$, we adopt a
fiducial value that is adjusted to reproduce the current climate of
Earth and is $93\%$ of that used by WK97: $D_{\rm fid} = 5.394\times
10^2 {\rm ~erg~cm^{-2}~s^{-1}~K^{-1}}$. For the diurnally-averaged
insolation flux $S$, we use the same standard formalism as WK97 (see
their Appendix).  Throughout this work, we adopt an Earth-like
planetary obliquity, $i=23.5$\degr, and assume zero eccentricity of
the planetary orbit for simplicity.

Equation~(\ref{eq:diffu eq1}) is solved on a grid uniformly spaced in
latitude with a highly efficient time-implicit numerical scheme and an
adaptive time--step, as described in \citet{hameury_et_al1998}. We impose
the boundary conditions $dT/d\lambda = 0$ at $\lambda = \pm 90\degr$
for symmetrical solutions. Resolution tests, which are further
described in Appendix~I, show excellent numerical convergence at our
fiducial resolution corresponding to $1.25\degr$ in latitude (145 grid
points pole--to--pole). All models are initiated with a uniform
temperature corresponding to an ice-free planet. To guarantee
satisfactory relaxation to a periodically-forced, seasonal climate
regime that is unaffected by details in initial conditions, all models
are integrated for at least 130 years, well in excess of the
thermal-inertial timescale of about a decade of the modeled
atmospheric layer. As shown in Appendix~II, specific choices of
initial conditions can potentially influence long-term model outcomes
by inducing a dynamical transition to a fully ice-covered (snowball)
climate even when ice-free initial conditions are specified. We
systematically initialize our models with temperatures well above
freezing (e.g., $T \gsim 300$~K for Earth-like planets), which
guarantees that partially ice-covered climates are favored over fully
ice-covered snowball states. This is similar to excluding ``cold
start'' scenarios in global radiative balance studies of habitability
(e.g., Kasting et al. 1993).

To summarize, we make the following assumptions in the models
presented here:
\begin{enumerate}
 \item {\it Heating/Cooling} -- The heating and cooling functions are
    given by the diurnally averaged insolation from a sun--like
    (1~$M_\sun$, 1~$L_\sun$) star, with albedo and insolation
    functions given in Table~\ref{tab:one}.
 \item {\it Latitudinal Heat Transport} -- We test the influence on
    climate of three different efficiencies of latitudinal heat
    transport, within the diffusion equation approximation: an
    Earth-like diffusion coefficient, and a diffusion coefficient
    scaled down and up by a factor of 9 (which correspond to 8-hour
    and 72-hour rotation according to the $D\propto {\Omega_p}^{-2}$
    scaling described in \S~\ref{sssec:Clim Stab} below).
 \item {\it Ocean Coverage} -- Every model has the same land--fraction
    in every latitude band.  For most of the models presented in this
    paper, each latitude band has an Earth--like 30\%:70\% land:ocean
    ratio.  We also present a series of models that represent a
    desert-world, with a 90\%:10\% land:ocean ratio.
 \item {\it Initial Conditions} -- The models all have a hot-start
    initial condition, with the planet's temperature uniform and at
    least 350~K, to minimize the chances of ending up in a global
    snowball state owing solely to the choice of initial conditions
    (see Appendix~II for more details).  Time begins at the northern
    winter solstice.
\end{enumerate}

\subsubsection{Model Limitations}
\label{sssec:limits}

Climate systems involve interactions over a wide range of time and
length scales, so that any given model (even a modern General
Circulation Model -- GCM) only emphasizes a limited range of
scales. This is particularly true of the simple 1D EBMs considered
here. Before presenting results from EBM calculations, it is therefore
important to clarify what the main limitations of these models are.

One type of limitation is related to the low spatial dimensionality of
our 1D EBM.  Our latitudinal EBM includes only a diurnally-averaged
formulation of insolation. While this is acceptable for a planet like
the Earth, with a relatively fast spin rate and significant overall
thermal inertia from a large ocean fraction, this simplification will
not be generally valid. In addition, since the models do not resolve
planetary longitudes, the surface temperature at a given latitude must
be interpreted as an average over the corresponding latitude
circle. This can lead to some ambiguities, in the sense that a
2-dimensional model resolving the longitudes would generally find
different physical conditions to be present over land vs. oceanic
regions. This land/ocean distinction, which is generally a function of
latitude, like on Earth, is not well addressed by our
longitudinally-averaged 1D EBMs.

A second type of limitation is related to the restricted range of
timescales captured by our models. Our work focuses on the atmospheric
response to a seasonal forcing and therefore emphasizes climate
features emerging typically on orbital timescales. As a result, it
omits many of the slower processes involved in determining the
long-term climate on a planet like the Earth. For example, oceans are
only treated as partial heat reservoirs in our models, thereby
neglecting their circulatory effects on longer timescales. Similarly,
the role of the carbonate--silicate cycle in regulating the
atmospheric CO$_2$ composition (e.g., \citealt{kasting_et_al1993};
WK97) and the resulting effects of atmospheric CO$_2$ on both
cloud--formation and the greenhouse effect are entirely neglected in
our EBMs.  In particular, on timescales longer than considered in our
models, it is expected that the type of CO$_2$ warming discussed by
\citet{kasting_et_al1993}, \citet{forget+pierrehumbert1997}, and
\citet{mischna_et_al2000} would extend the outer boundary of the
habitable zone.  A massive release of CO$_2$ in Earth's atmosphere,
via an asymmetric carbon-silicate cycle for millions of years, is also
the leading candidate scenarios invoked to explain how the
paleoclimate was ultimately able to escape a globally-frozen snowball
state \citep{hoffman+schrag2002}. On the much shorter timescales
described by our models, however, snowball events are simply
semi-permanent states of the climate.

Finally, we note that there are numerous additional atmospheric
processes that are ignored from our modeling strategy. They include,
for instance, feedback effects from clouds on the albedo and
IR-cooling functions or the possibility of transient oceanic
evaporation and partial atmospheric escape in the regionally hottest
models considered. Despite these important shortcomings, our EBMs are
useful to explore many important dynamical climate features possibly
relevant to planetary habitability.

\subsubsection{Model Validation}
\label{sssec:valid}

In what follows, {the model we refer to as the ``fiducial'' one is the
Earth--like EBM} built from the IR-cooling and albedo functions listed
as model~2 in Table~\ref{tab:one}, using a uniform $70$\% ocean
fraction, a latitudial heat diffusion constant $D=D_{\rm fid}$ and
``hot-start'' (ice-free) initial conditions.  We find very good
agreement between the latitudinal surface temperature distribution
predicted by this fiducial model and the one observed on Earth.  The
National Center for Environmental Prediction (NCEP), in conjunction
with the National Center for Atmospheric Research (NCAR), has released
6--hourly global temperature maps for Earth that are the results of a
detailed global climate model that is tightly constrained by observed
temperatures \citep{kistler_et_al1999, kalnay_et_al1996}.  For model
validation, we use these data for model validation and the
phenomenalogical fit to Earth's annual mean latitudinal temperature
profile proposed by \citet{north+coakley1979}:
\begin{equation}
\label{eq:north+coakley1979}
T[\lambda] = 302.3\,{\rm K}- 45.3 \sin^2[\lambda].
\end{equation}

\begin{figure}[h!]
\plotone
{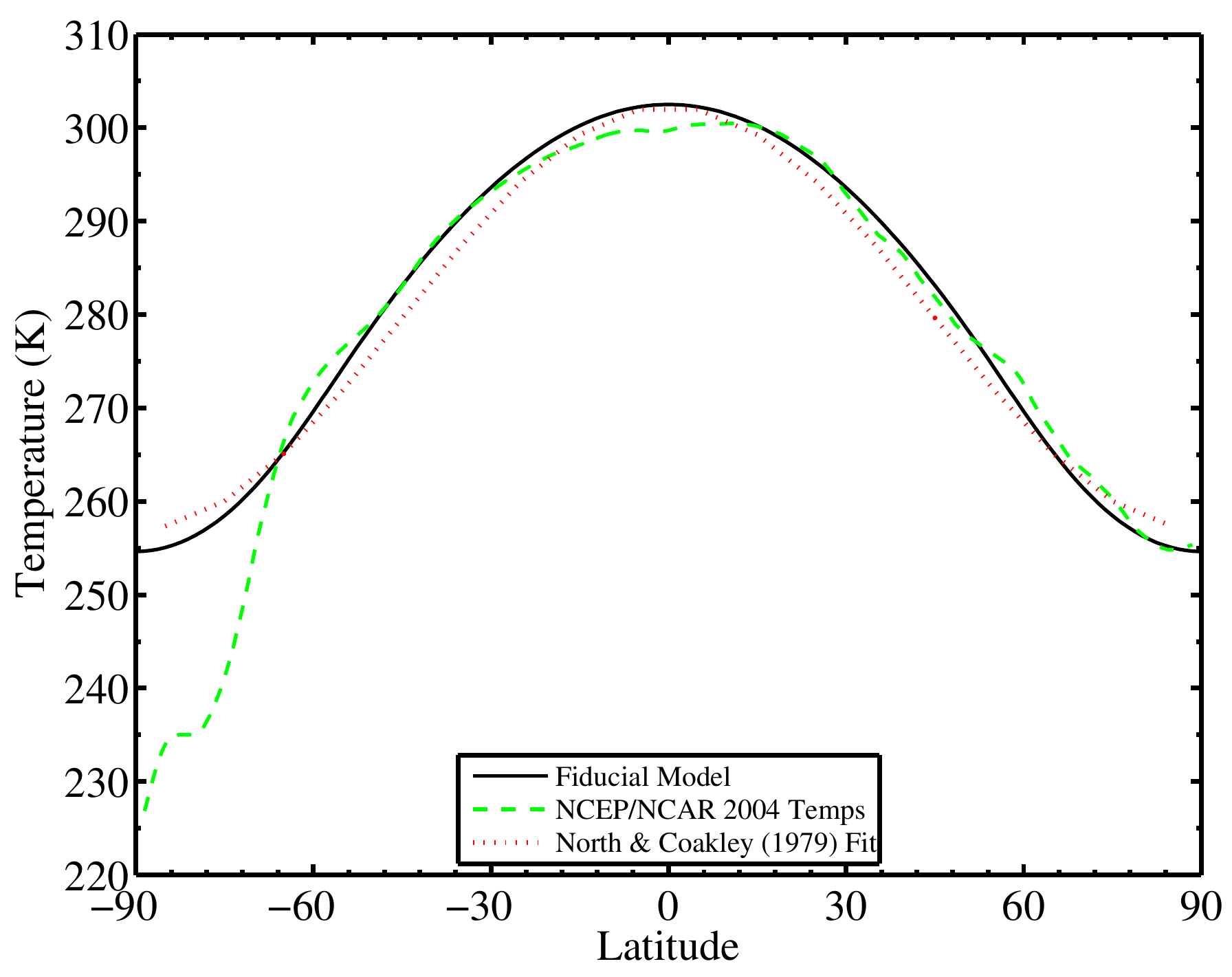}
\caption{Model validation based on the mean latitudinal temperature
profile of the Earth.  The solid line shows the annual mean
temperature profile in our fiducial Earth-like model. The dashed line
shows the average taken from the NCEP/NCAR global temperature data in
2004. The dotted line shows the fit to Earth's mean temperature
profile proposed by \citet{north+coakley1979}. The discrepancy at the
South Pole is due to Antartica.}
\label{fig:us+NC1979}
\end{figure}

Figure~\ref{fig:us+NC1979} shows the annual mean temperature
distribution in our fiducial model (solid line) after $130$ years of
thermal relaxation. It is within 5~K of the phenomenalogical fit in
Eq.~(\ref{eq:north+coakley1979}), shown as a dotted line, at all
latitudes.  The model also reproduces quite well the zonally and
temporally averaged surface temperatures from NCEP/NCAR for the year
2004\footnote{The match is good for other years as well, but only the
2004 data product is shown here.} at all latitudes north of
$60\degr$~S. The North--South asymmetry of the land--distribution on
Earth, due to the presence of a continental circumpolar region --
Antarctica -- in the south, breaks the symmetry of the temperature
profile.\footnote{The figure illustrates the potential magnitude of
effects due specific land/ocean configurations.}  It is therefore not
surprising that our simple, symmetrical EBM fails to reproduce the
sharply colder temperatures near the South Pole.

\begin{figure}[h!]
\plotone
{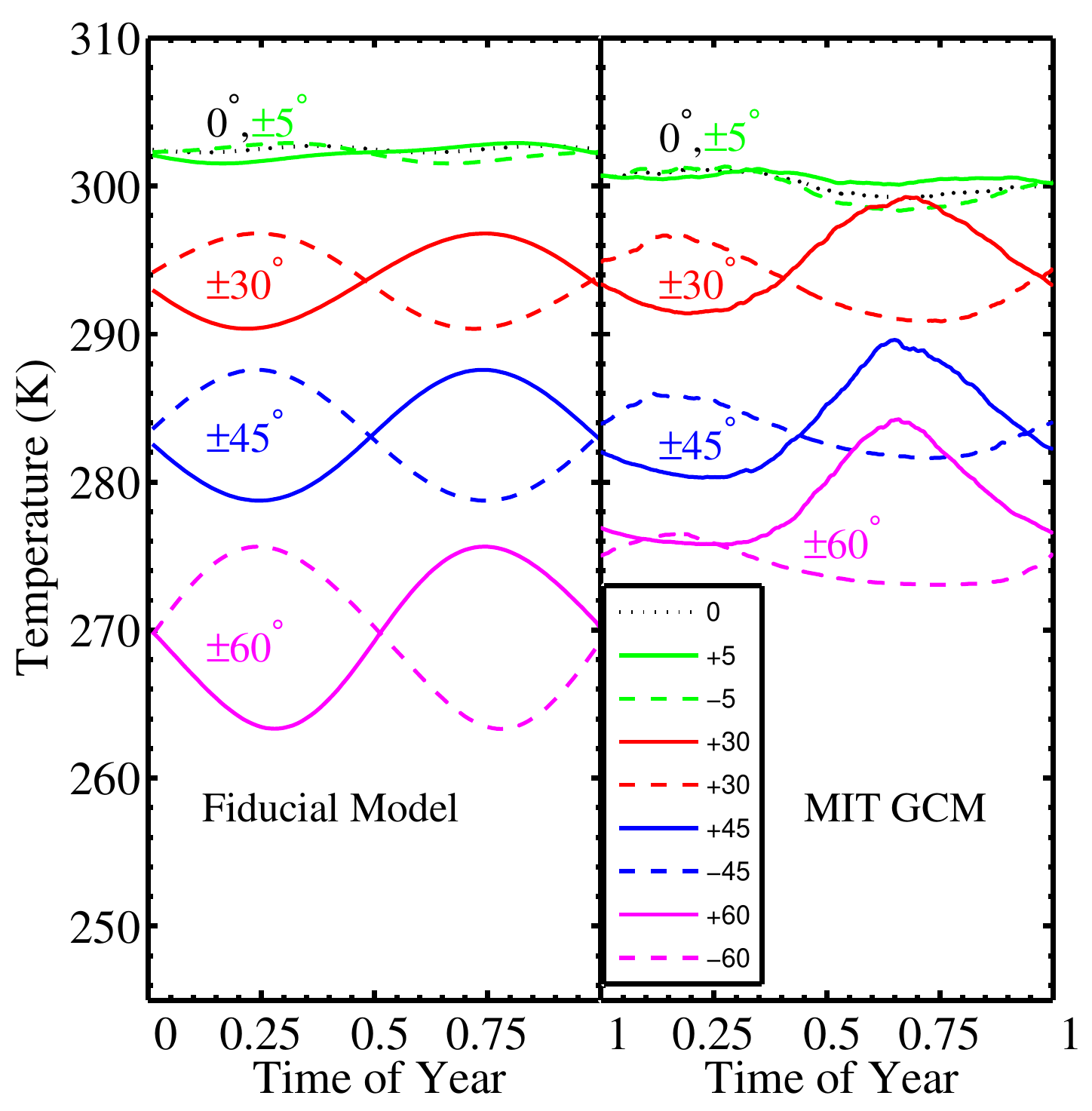}
\caption{Model validation based on seasonality at different latitudes.
The left panel shows temperatures as a function of time for several
latitudes in our fiducial Earth-like model.  The right panel shows the
2004 ocean--surface temperatures in the MIT Oceanic GCM
\citep{wunsch+heimbach2007} for the same latitudes.  In both panels,
solid lines correspond to temperatures in the North and dashed lines
to temperatures in the South.  Time of year is measured in fraction of
a year from the northern winter solstice.  Note that our model is
North--South symmetrical while the MIT Oceanic GCM is not.  Extrema of
temperature occur approximately one quarter cycle after the extrema in
forcing, as expected for a seasonally-forced planet in the limit of
large thermal inertia \citep[e.g.,][]{north+coakley1979}.}
\label{fig:us+MIT}
\end{figure}

In addition to reproducing annual mean properties, the fiducial model
also captures important seasonal variations seen in temperature data
from the Earth or in advanced GCMs of the Earth.  In particular, the
temperature response is found to lag the solar forcing in our model by
about a quarter annual cycle: the warmest temperatures happen at the
local autumnal equinox and the coolest temperatuers at the local vernal
equinox.  This well known phase lag is the result of the high heat
capacity of the wind-mixed ocean layer (e.g.,
\citet{north+coakley1979}), which covers 70\% of our model Earth-like
planet.  For detailed comparison, we obtained surface layer outputs
from the full three--dimensional MIT Oceanic GCM
\citep{wunsch+heimbach2007}.  Figure~\ref{fig:us+MIT} shows the
temperature as a function of time of year at latitudes from the
equator to $\pm60\degr$, in our model and the uppermost ocean layer of
the MIT GCM (5~m depth) for the year 2004.  Time of year is measured
in fraction of a year from the northern winter solstice.  Our model is
symmetrical with respect to the equator, so there is no real
distinction between North and South, but the MIT GCM is not.  Both
models show the same response--lag of $\sim 90\degr$.  Towards the
poles, our model gets significantly colder than the (ocean only) GCM,
but at equatorial and mid--latitudes, remarkably, the deviations are
in most places significantly less than 5~K.

Although our fiducial model is quite successful in reproducing various
seasonal features of Earth's temperature distribution, we caution that
this alone does not guarantee that predictions will remain accurate
for parameters other than the fiducial ones (e.g., for changes in
orbital distance leading to systematically hotter or colder climates).
To achieve a reasonably good fit to the Earth, various paramaters in
the model were adjusted (e.g., constants in model~2 of
Table~\ref{tab:one} or value of $D_{\rm fid}$).  Since the model is
based on parameterizations, rather than fundamental laws, it is
possible that its behavior will deviate quantitatively from what
Earth's response would be when new regimes of climate are
explored. Nevertheless, the detailed model validation performed here
with the Earth and the physics-based prescriptions adopted in our
modeling strategy should guarantee that our EBM predictions are
qualitatively robust. As we shall see below, this notion is supported
by the generally good agreement of results obtained from the three
different sets of atmospheric IR-cooling and albedo models listed in
Table~\ref{tab:one}.

\section{Study of Habitability}
\label{sec:Hab Zones}

Equipped with a well-tested and validated EBM, we now reconsider
several issues related to the habitability of Earth-like planets. In
the present study, we emphasize differences between EBM and global
radiative balance calculations, focusing on the concept of fractional
habitability, and we highlight the potentially important role played
by climate stability against snowball events in determining the
habitability of an Earth-like planet.

\subsection{Climate Dynamics and Seasonality}
\label{ssec:Temp}

\subsubsection{Climate Stability}
\label{sssec:Clim Stab}

Even a simple EBM such as ours can exhibit a surprisingly rich climate
phenomenology, as compared to the outcomes of global radiative balance
models. A particularly striking example of this phenomenology is the
susceptibility of the climate system to albedo feedback effects that
can induce transitions from partially ice--covered states to
globally-frozen snowball states. In the context of habitability
studies, these transitions are important because they could render a
planet essentially ``uninhabitable,'' unless transitions back to
partially ice--covered or ice--free states are possible. Since the
seminal work of Budyko (1969) and Sellers (1969), it has been well
known that Earth's partially ice-covered climate allows for
transitions to globally-frozen snowball states in response to
relatively minor changes in forcings (e.g.,
\citet{hartmann1994,ghil2002,hoffman+schrag2002}).

Climate stability is a subtle issue, even in the context of a 1D
diffusion model. \citet{held_et_al1981} describe how it is ultimately
an interplay between latitudinal heat transport and albedo feedback
effects that determines climate stability.  Here, we will not attempt
a thorough exploration of climate stability with our EBM but instead
will illustrate how this issue could be critical in determining the
habitability of a seasonally-forced Earth-like planet.

Latitudinal heat transport by atmospheric motions is inhibited by
Coriolis effects on a rotating planet (e.g.,
\citet{pedlosky1982,holton1992}). As a result, latitudinal heat
transport should be typically reduced on a pseudo-Earth rotating
faster than the current Earth. There is currently no first principle
theory allowing reliable estimates of how much transport would be
reduced on a faster rotating Earth-like planet. There have been,
however, simple scaling arguments proposed in the geophysical
literature.  Here, in the interest of simplicity, we will use the
``thermal Rossby number'' scaling advocated by \citet{farrell1990},
which suggests that $D$ is $\propto {\Omega_p}^{-2}$, where $\Omega_p$
is the angular rotation rate of the planet (see also
\citealt{stone1973}). We note that detailed GCM experiments for
slow-rotating Earths do not necessarily support the extreme simplicity
of the above scaling \citep{delgenio_et_al1993,delgenio+zhou1996}.

Figure~\ref{fig:temp maps} shows space-time diagrams of temperature in
two Earth-like models differing only in the magnitude of latitudinal
heat transport.  The top two panels show temperatures for the full 130
years (upper left) and for the last 5 years (upper right) of our
fiducial Earth-like model.  The bottom two panels show temperatures,
over the same time spans, in a model of a pseudo-Earth that is in all
respects identical to our fiducial model, except for a diffusion
coefficient $D$ reduced by a factor of 9 (which, according to the
above scaling, corresponds to a planet rotating three times as fast as
Earth does\footnote{We note that it has been suggested that the young
Earth may have spun four times as fast as it currently does
\citep{canup2004}.}).  Both models start with a uniform $T=350$~K,
but their evolutions follow strikingly different trajectories.  The
fiducial model settles to the standard, partially ice-covered
solution.  In contrast, the model with inefficient latitudinal heat
transport undergoes rapid evolution through the first $\sim 50$ years,
and then abruptly, over the course of a single year, transitions to a
permanent snowball state with a mean temperature around 200~K.

Note that in the Earth--like fiducial model (upper right panel), the
poles are, unsurprisingly, the coldest parts of the planet.  By
contrast, in the fast-Earth model (lower right panel), the poles
become nearly as warm during local summer as the warmest parts of the
planet.  This is because, with our albedo function (model~2 in
Table~\ref{tab:one}), the entire planet has identical albedo in the
snowball state, and so the poles, which actually experience greater
diurnally averaged insolation during local summer than any other part
of the planet, can warm a great deal in response to the high solar
irradiance.  In the fiducial Earth-like model, on the other hand,
parts of the planet that are above freezing absorb significantly more
solar radiation, owing to their lesser albedo, than parts that are
well below freezing; this is what keeps the poles relatively cold
year--round in the fiducial model.

The main conclusion to be drawn from this comparison between our
fiducial Earth-like model and a fast-rotating pseudo-Earth model is
that issues of climate dynamics, especially climate stability with
respect to snowball transitions, can critically influence the
habitability of Pseudo--Earth planets. While the planetary rotation
rate is simply inconsequential for global radiative balance models of
habitability, it becomes a potentially important planetary attribute
for habitability in more elaborate dynamical climate models, even for
Earth-like planets at $1$~AU from a Sun-like star, as was assumed in
both models shown in Figure~\ref{fig:temp maps}.

\subsubsection{Comparison to Global Radiative Balance Results}

We now relax our $1$~AU assumption and calculate EBMs for Earth-like
planets at a range of orbital distances from their Sun-like star. We
place our model planets at $41$ distances from the star, in the range
$0.3$ to $1.3$~AU, with $0.025$~AU steps.  Each model is run for $130$
local years (orbits), which is sufficient for adequate model
relaxation from the hot-start to the final periodic climate solution.
At every distance, we also calculate the radiative equilibrium
temperature that obtains from the global average insolation, so that
we can directly compare our time-dependent EBM results to those of an
equivalent global radiative balance model.

\begin{figure}[h!]
\plotthree
{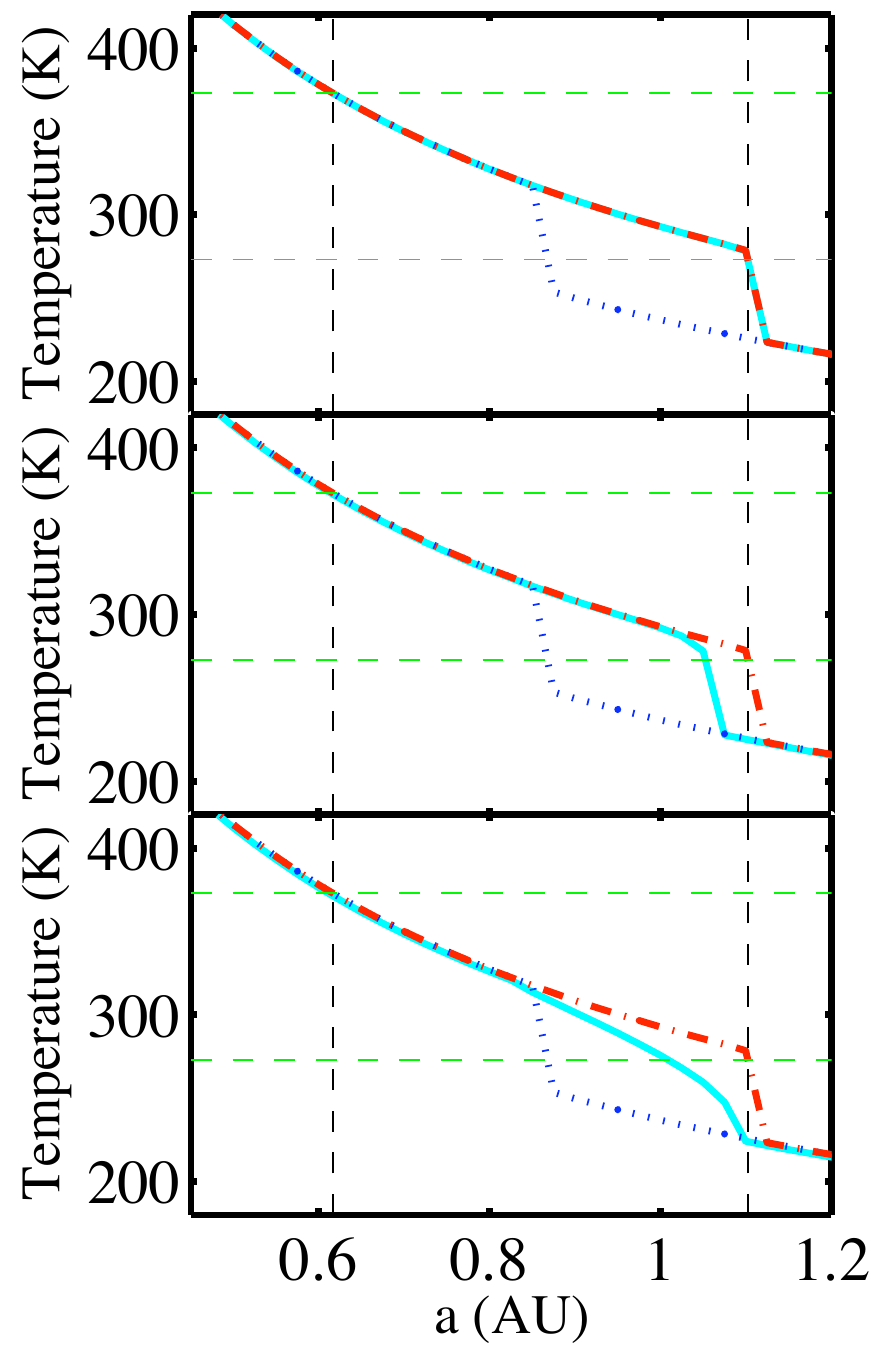}
{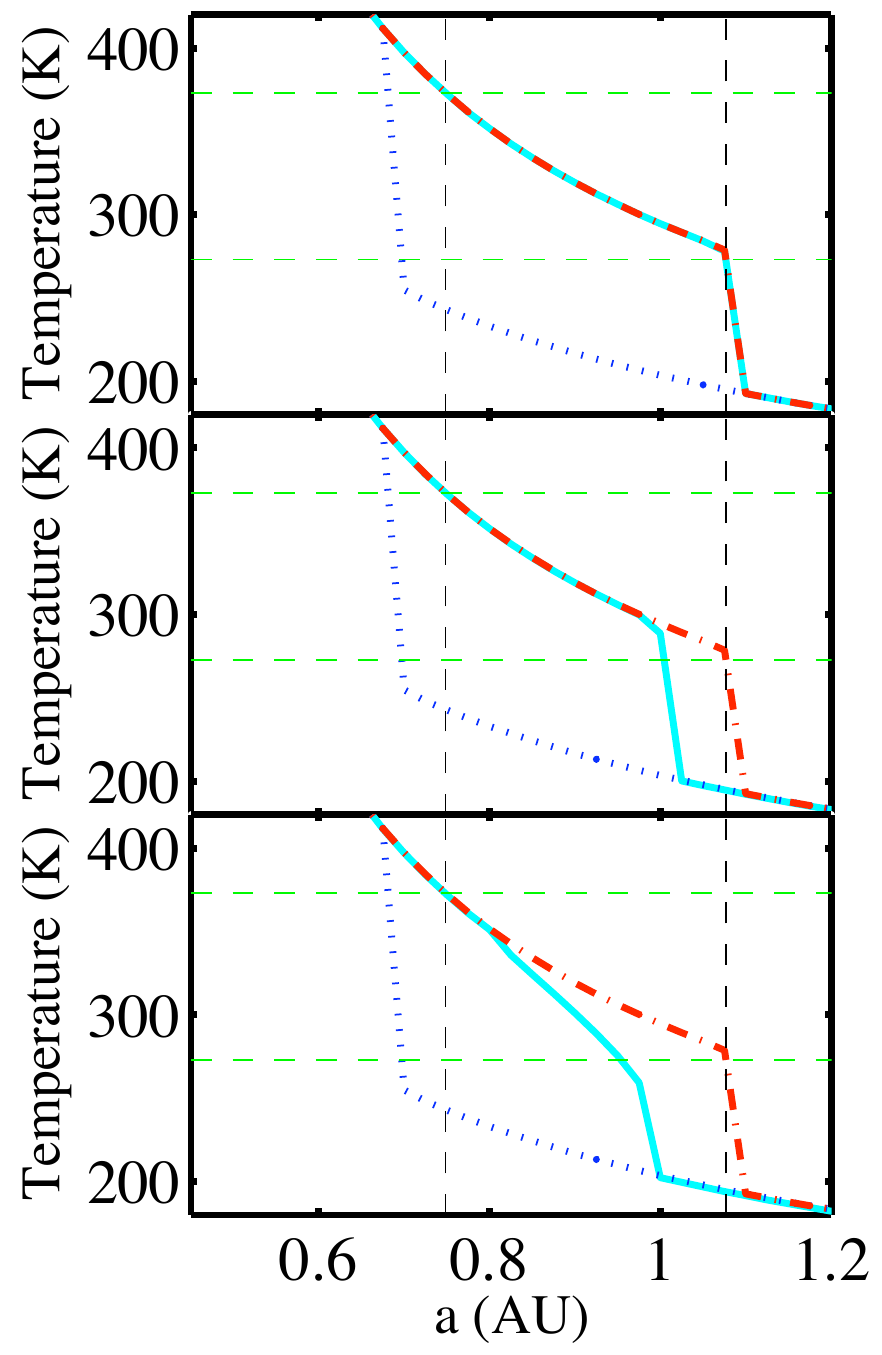}
{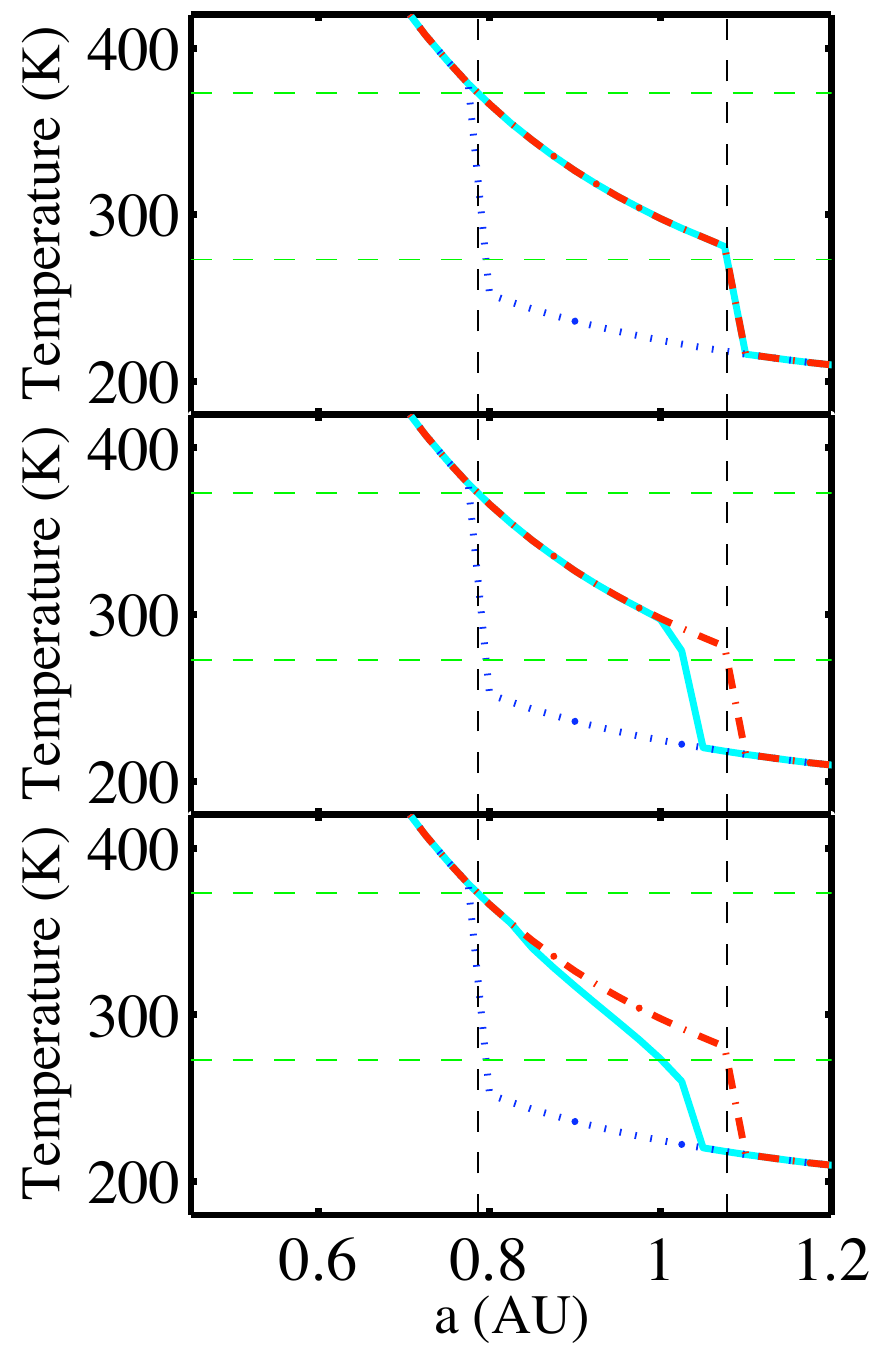}
\caption{Comparison between global radiative equilibrium temperatures
and averaged temperatures in our seasonal EBM, for nine different
pseudo-Earth models. From left to right, the three columns show
results obtained using models~1, 2 and~3 in Table~\ref{tab:one} for
the IR-cooling and albedo functions, respectively. In each column,
from top to bottom, panels correspond to values $D=9\,D_{\rm fid}$,
$D_{\rm fid}$, and $D_{\rm fid}/9$ for the latitudinal diffusion
coefficient used in the EBM.  In each panel, the solid curve shows the
global and annual mean temperature in the EBM over a range of orbital
distances. The dotted and dashed-dotted curves show the corresponding
low- and high-temperature radiative equilibrium solutions,
respectively.  The mean temperature in the EBM deviates from the
high-temperature radiative equilibrium solution for small and moderate
values of the coefficient of latitudinal heat diffusion.}
\label{fig:temp}
\end{figure}

As discussed in \S~\ref{ssec:globrad}, there is in general a range of
orbital distances around $\sim 1$~AU for which multiple radiative
equilibrium solutions exist.  We focus exclusively on the two stable,
low and high temperature solutions here. Each panel in
Figure~\ref{fig:temp} compares the radiative equilibrium solutions
(dotted and dashed-dotted curves) to results for the annual mean
temperature weighted by surface area in our EBM calculations (solid
line), for a range of orbital distances.  From left to right, the
three columns show results using models~1, 2, and~3 in
Table~\ref{tab:one}, respectively, for the IR-cooling and albedo
functions. In each column, EBM results are shown for three
efficiencies of latitudinal heat transport, corresponding to
$D=9\,D_{\rm fid}$, $D_{\rm fid}$, and $D_{\rm fid}/9$, from top to
bottom. Our fiducial Earth-like model corresponds to the central
panel.

At small and large orbital separations, the EBM and radiative-balance
curves nearly coincide with each other, indicating that the unique
radiative equilibrium solution is very close to the averaged
temperature in our seasonal EBM. At intermediate separations
corresponding to partially ice-covered climates, however, the curves
separate. The distinction among the three curves is particularly
important for EBMs with weak latitudinal heat transport (bottom
panels) because EBMs with efficient latitudinal heat transport (top
panels) have generally more uniform temperature distributions, which
is close to the implicit averaging assumption made in global radiative
balance models.

Each panel in Figure~\ref{fig:temp} has horizontal dashed lines at
273~K (water freezing) and at 373~K (water boiling) and vertical
dashed lines indicating where the high temperature radiative
equilibrium solution crosses these 273~K and 373~K limits.  In other
words, the vertical dashed lines indicate the orbital extent of the
radiative equilibrium habitable zone (hereafter, REHZ) for globally
averaged conditions, analogous to the calculations in
\citet{kasting_et_al1993}.  The extent of the habitable zone defined
on the basis of the globally and annually averaged temperature in our
EBMs is the range of orbital radii, $a$, for which the solid curve
remains between the two horizontal dashed lines (indicating surface
liquid water conditions for an Earth-like planet). In all the models
considered here, this EBM-average habitable zone tends to have nearly
the same inner boundary as the REHZ, but its outer boundary can be
significantly closer to the Sun-like star than that of the
REHZ. Indeed, in EBMs with fiducial or reduced efficiencies of
latitudinal heat transport, the climate typically makes dynamical
transitions to a fully ice-covered snowball state at orbital distances
well inside of what global radiative balance models might
indicate. Although quantitative discrepancies exist, this qualitative
behavior is common to all three models with different IR-cooling and
albedo functions shown in Fig.~\ref{fig:temp}. Note that the dynamical
transition to a snowball state shown in Fig.~\ref{fig:temp maps} for a
fast-rotating pseudo-Earth at $1$~AU is included in the lower middle
panel of Fig.~\ref{fig:temp}.

While useful for comparisons between EBM and global radiative balance
results, the diagrams shown in Fig.~\ref{fig:temp} do not accurately
capture the range of surface temperature conditions found in our
seasonal EBMs. Indeed, for every globally averaged annual mean
temperature from an EBM, there must exist a range of temperatures
above and below this average, both in a regional and a temporal
sense. In general, planets can therefore be habitable beyond what
globally averaged studies might indicate. This leads us to give a
formal definition of the notion of fractional habitability.

\subsection{Fractional Habitability}
\label{ssec:FracHab}

For a fixed set of forcing and response parameters, a planet might be
habitable (i.e., have a surface temperature between 273~K and 373~K)
over only a portion of its total surface area or for only a fraction
of its orbit.  To quantify the various possible outcomes in our EBMs,
we develop several metrics of fractional habitability.

Let $H[a,\lambda,t]$ be the ``habitability function'', equal to 1 if
latitude $\lambda$, on a model planet at a distance $a$ from its star,
has a habitable temperature at time $t$, and 0 otherwise:
\begin{equation}
H[a,\lambda,t] = \left\{ \begin{array}{ll}
1 & {\rm if}\, 273 \leq T[a,\lambda,t] \leq 373, \\
0 & {\rm otherwise.}
\end{array} \right. .
\label{eq:hab_func}
\end{equation}
Using $H$, we define three habitability fractions.

At each latitude, we calculate the fraction of the year that that
portion of the planet spends in the habitable temperature range,
\begin{equation}
\label{eq:f_time}f_{\rm time}[a,\lambda] = \frac{\int_{t=0}^{P}H[a,\lambda,t] dt}{P},
\end{equation}
where $f_{\rm time}$ is the habitable fraction as a function of
latitude, and $P$ is the length of the year at orbital separation $a$.
At each time of year, we calculate the fraction of the planet's
surface area that is habitable:
\begin{equation}
\label{eq:f_lat}f_{\rm area}[a,t] = \frac{\int_{\lambda=-\pi/2}^{\pi/2}H[a,\lambda,t] \cos[\lambda] d\lambda}{2}.
\end{equation}
Finally, at each orbital separation, we calculate the net fractional
habitability, which is the area--weighted fraction of the $\lambda-t$
plane over which the planet is habitable:
\begin{equation}
\label{eq:f_hab}f_{\rm hab}[a] = \frac{\int_{\lambda=-\pi/2}^{\pi/2} \left\{ \int_{t=0}^{P}H[a,\lambda,t] dt \right\} \cos[\lambda] d\lambda }{2 P}.
\end{equation}

\subsubsection{The Case of Earth}
\label{sssec:comp}

\begin{figure}[h!]
\plotone {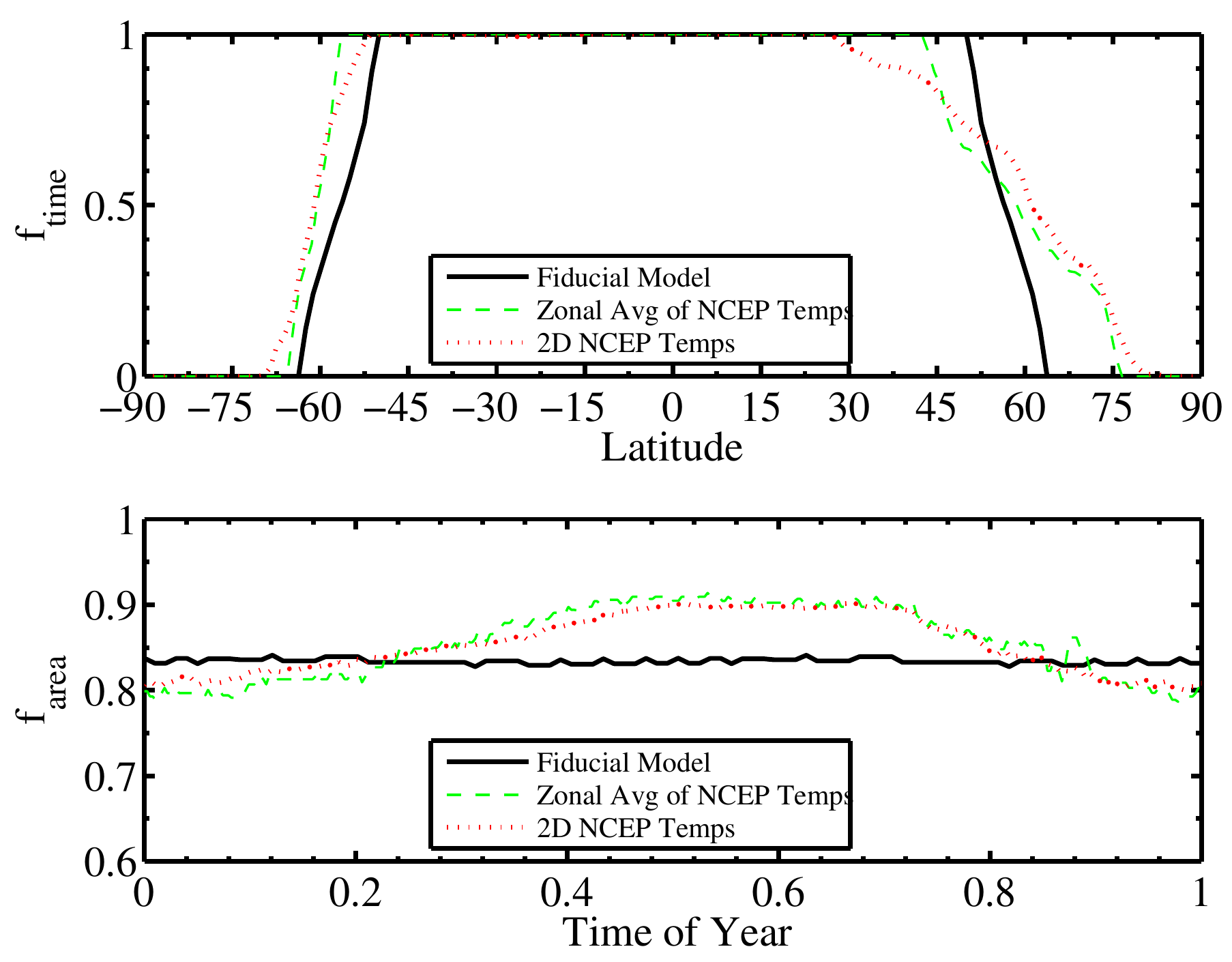}
\caption{Temporal and regional habitability fractions of the Earth and
in our fiducial Earth-like model.  The top panel shows $f_{\rm time}$,
the fraction of the time that each latitude band spends between 273~K
and 373~K.  The bottom panel shows $f_{\rm area}$, the fraction of the
land area that is between 273~K and 373~K at different times of year,
measured from the Northern winter solstice.  In both panels, the solid
line shows results for our fiducial model, the dashed line shows
results for the Earth using the zonally-averaged NCEP/NCAR temperature
data in 2004, and the dotted line shows results for the Earth using
the full two--dimensional NCEP/NCAR 2004 temperature data.}
\label{fig:hab_comp}
\end{figure}

Before analyzing the fractional habitability of model planets with
different characteristics than Earth, it is instructive to consider
the fractional habitability of Earth itself and of our fiducial
Earth-like model.  We return to the NCEP/NCAR temperature data in 2004
for this comparison.  Figure~\ref{fig:hab_comp} shows a remarkably
good agreement between the temporal and regional habitability of our
fiducial model (solid line) and that of Earth (dashed and dotted
lines).  We analyzed the 2004 Earth temperatures in two different
ways.  In order to create a variable that is directly comparable to
our (one dimensional) model temperatures, we must perform zonal
averages.  In Figure~\ref{fig:hab_comp}, statistics derived from first
zonally-averaged temperatures are represented by dashed lines.  We
also used the entire data set (latitude, longitude and time) to
calculate habitability fractions before performing zonal averages.
Statistics derived from this full set of temperature data are
represented by dotted lines.

The top panel in Figure~\ref{fig:hab_comp} shows the fraction of the
time that each latitude strip spends at or above 273~K.  There are
some differences between the dashed and the dotted lines, most
noticeably in the steepness of the descent from 100\% habitable to 0\%
habitable moving from equator to North Pole.  The crucial feature of
this plot, from a model validation perspective, is that the $f_{\rm
time}$ values derived from zonally averaged temperatures on Earth have
a very steep descent in both hemispheres, similar to what is seen in
our fiducial model.

The bottom panel in Figure~\ref{fig:hab_comp} shows the fraction of
the surface area that is habitable as a function of time of year
(measured from the Northern winter solstice).  The fiducial model
maintains nearly constant regional habitability throughout the year.
Earth's regional habitability varies slightly over the year,
increasing a bit in the Northern summer because the North Polar ice
cap shrinks much more during its summer than the South Polar cap does
during its.

Finally, the net fractional habitability ($f_{\rm hab}$) of Earth and
of our fiducial model are just the average values of the curves shown
in the bottom panel of Figure~\ref{fig:hab_comp}.  Our fiducial model
is 83\% habitable. Earth was 85\% habitable in 2004 when calculated
from two--dimensional temperature data (this fraction varies by less
than 0.5\% from year to year). The net fractional habitability of
Earth was 86\% in 2004 when calculated from a temperature field that
is first zonally averaged.

\subsubsection{Temporal Habitability of Pseudo-Earths}
\label{sssec:f_time}

\begin{figure}[h!]
\plotthree {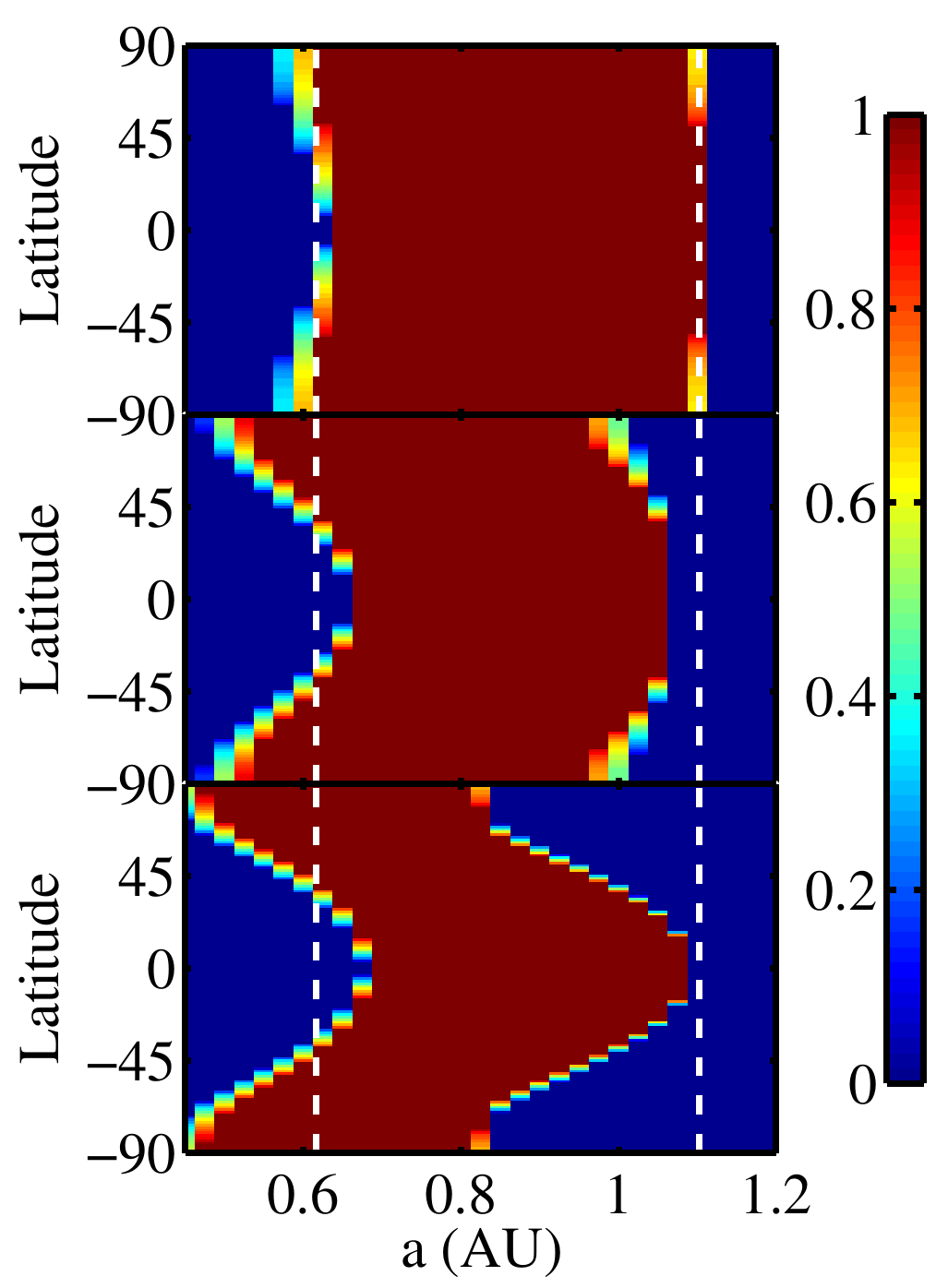} {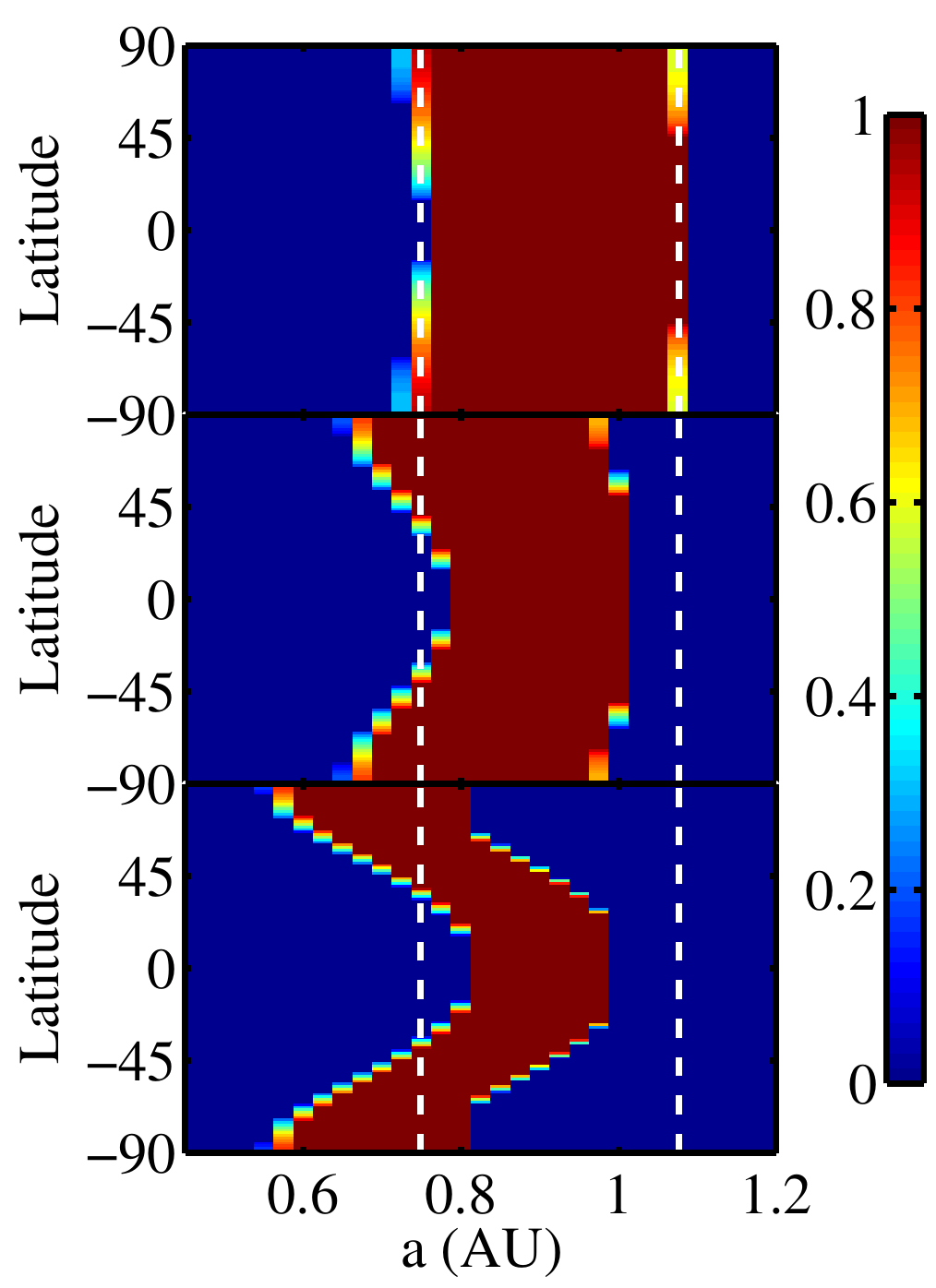} {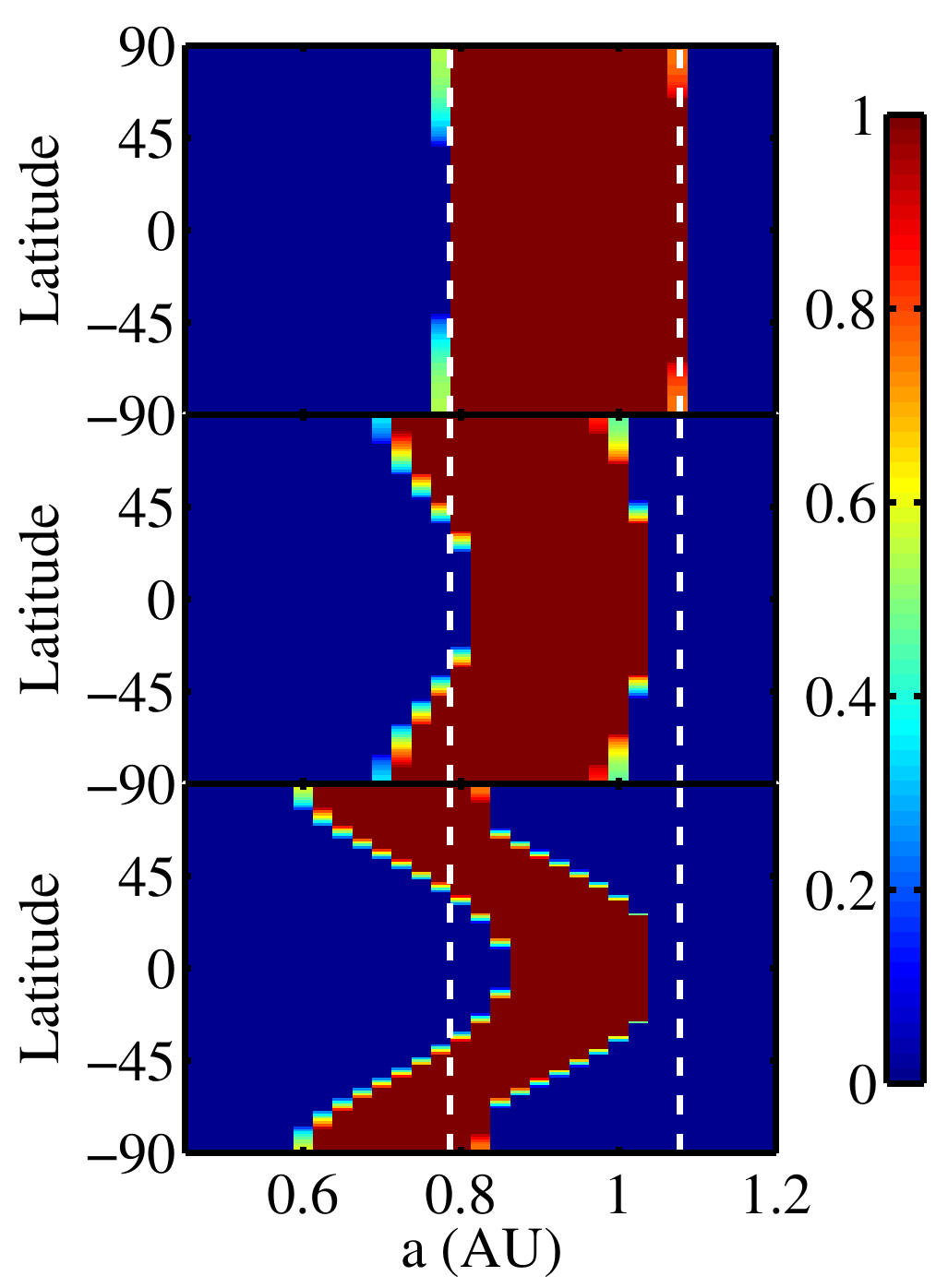}
\caption{Temporal habitability fraction, $f_{\rm time}$, as a function
of orbital distance and latitude. Results are shown for the same nine
pseudo-Earth models as in Fig.~\ref{fig:temp}, with the same panel
structure.  Vertical dashed lines indicate the orbital extent of the
radiative equilibrium habitable zone.  Fractional habitability
increasingly depends on latitude, and deviates from radiative
equilibrium results, when the efficiency of latitudinal heat transport
($D$) is reduced (from top to bottom).}
\label{fig:time hab}
\end{figure}

Figure~\ref{fig:time hab} presents a striking view of how regionality
depends on the efficiency of latitudinal heat transport with the same
nine pseudo-Earth EBMs as shown in Fig.~\ref{fig:temp}.  In this
figure, contours indicate the value of the temporal habitability,
$f_{\rm time}$, as a function of orbital separation ($a$) and latitude
($\lambda$).  When latitudinal heat transport is efficient -- in the
top panels -- temporal habitability depends little on latitude.  When
transport is inefficient -- in the bottom panels -- there is a great
deal of latitudinal dependence.  Since the poles tend to be colder at
a given separation, they tend to be habitable when the planet is
closer to the star.  For the same reason, the equator tends to be
habitable when the planet is further away from the star.

In each panel, the dashed vertical lines indicate the orbital extent
of the radiative equilibrium habitable zone (REHZ), taken from
Figure~\ref{fig:temp}.  In all models, allowing for regionality
extends the inner boundary of the habitable zone, relative to the
REHZ.  Models with low efficiencies of latitudinal heat transport tend
to have outer boundaries of the habitable zone that are closer to the
star than models with high $D$ values.  In essentially all cases, this
outer boundary is determined by a dynamical climate transition to a
globally-frozen snowball state. As a result, the relationship between
the outer boundary of the regionally habitable zone and $D$ is not
necessarily monotonic.  In the left column of Figure~\ref{fig:time
hab}, for example, the outer boundary of the regionally habitable zone
is minimum for $D=D_{\rm fid}$.  In all models, at a given latitude,
planets tend to be habitable either 100\% of the time or 0\% of the
time, with little space for intermediate cases.  As we shall see
below, this is a consequence of the assumed large ocean fraction
(70\%) on the model planets. The large resulting thermal inertia
minimizes the ability of a latitude band to swing below and above the
freezing point during an annual cycle.

One may ask what is the physical process that actually drives a model
planet to freeze over into a snowball state at a given orbital
separation.  For instance, in the central panel of
Figure~\ref{fig:time hab}, which shows results for our fiducial
Earth-like model, the planet has an Earth--like climate at $a=1$~AU
but is globally-frozen at $a=1.025$~AU.  What is different at 1.025~AU
than at 1.000~AU to induce such a dynamical transition?  Two
possibilities spring to mind: the reduced insolation or the longer
winters.  In order to distinguish between these two possibilities, we
ran two additional models: one with Earth--like insolation and a year
longer by a factor of $1.025^{3/2}$ (which did not freeze), and one
with insolation reduced by a factor of $1.025^2$ and a year the length
of Earth's (which did freeze).  This indicates that the reduced
insolation is the dominant effect leading to snowball transitions in
this model. As shown explicitly in Appendix~II, however,
longer winters can also contribute to dynamical snowball transitions.

\subsubsection{Regional Habitability of Pseudo-Earths}
\label{sssec:f_lat}

\begin{figure}[h!]
\plotthree {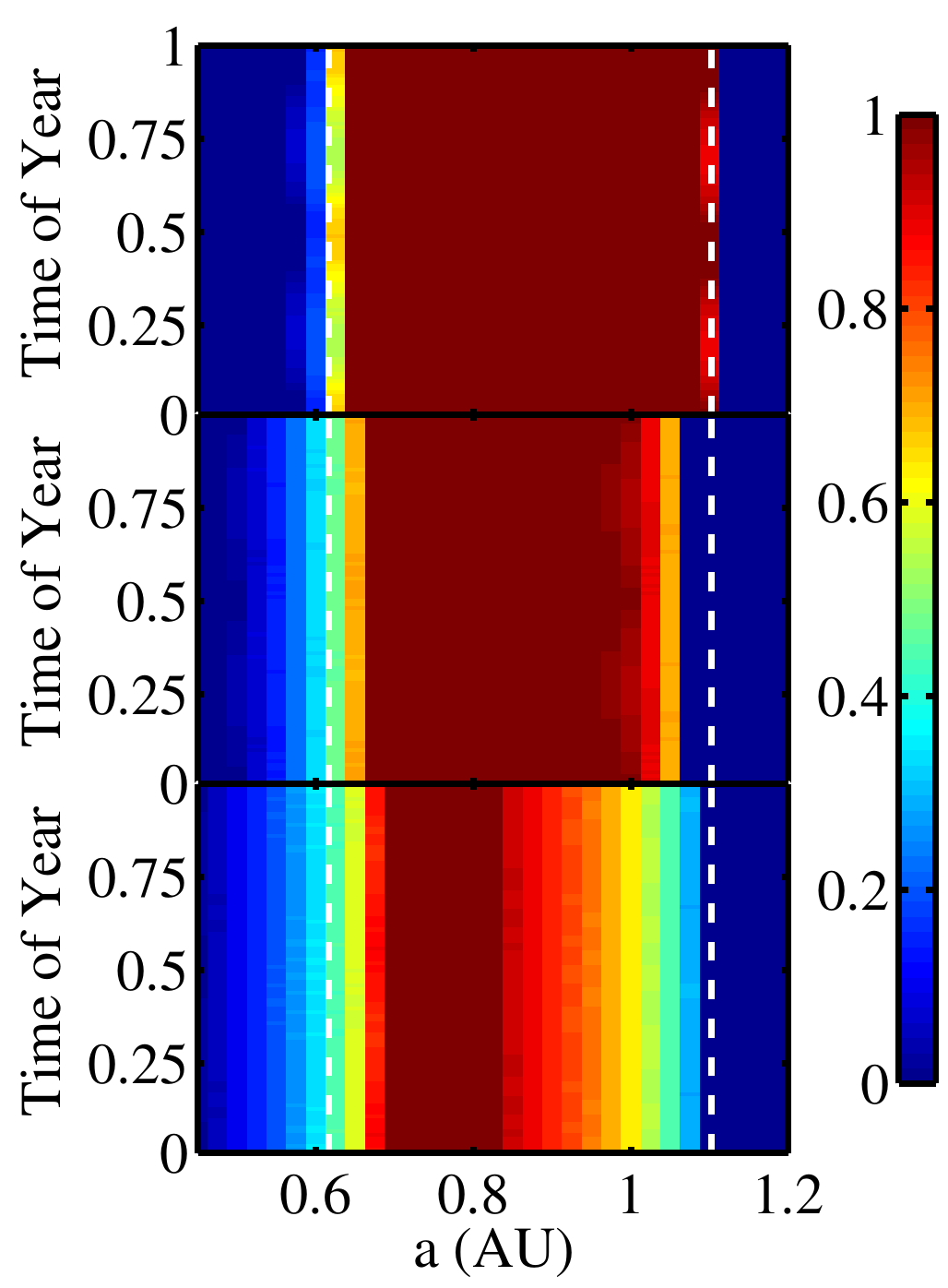} {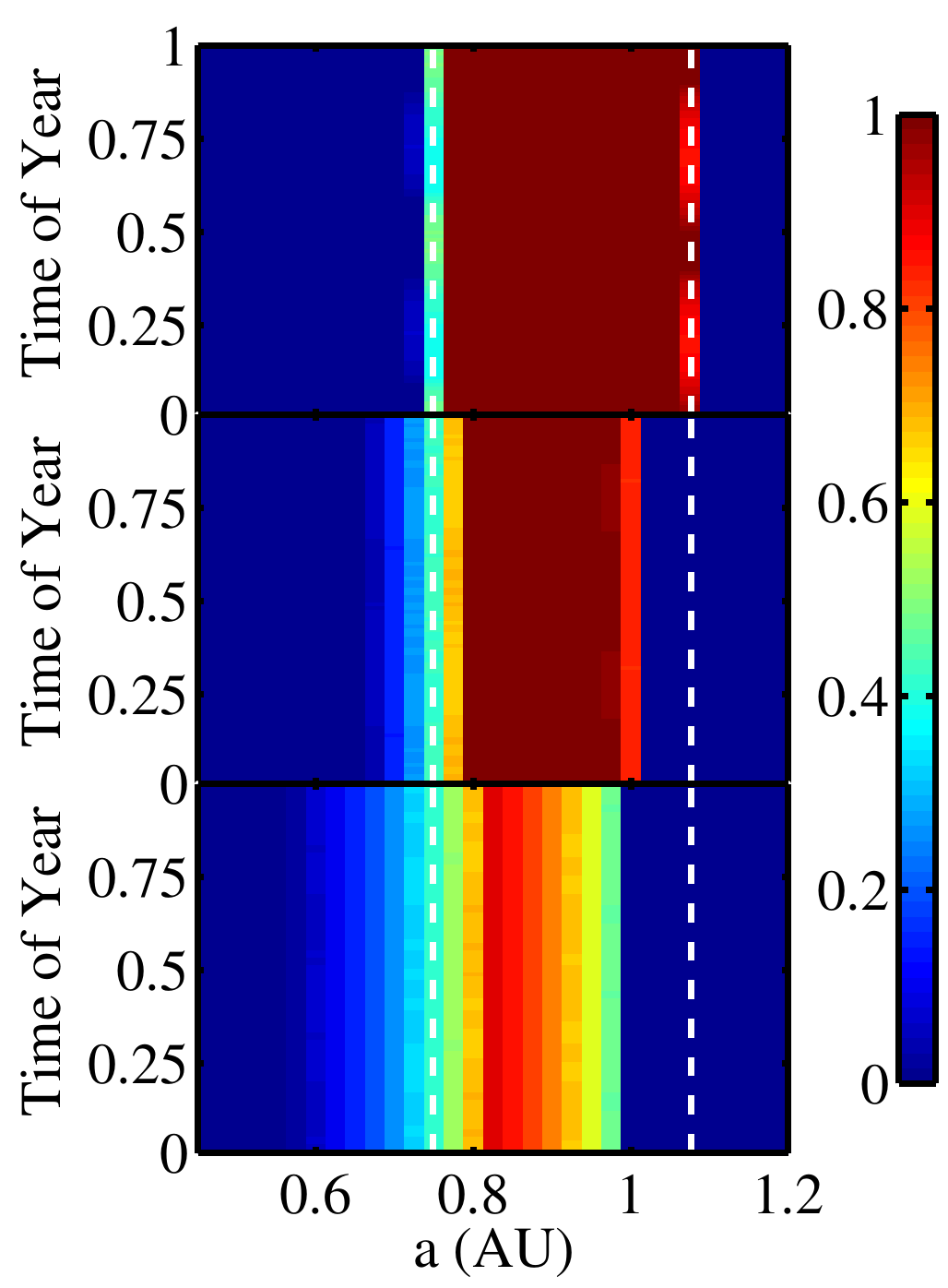} {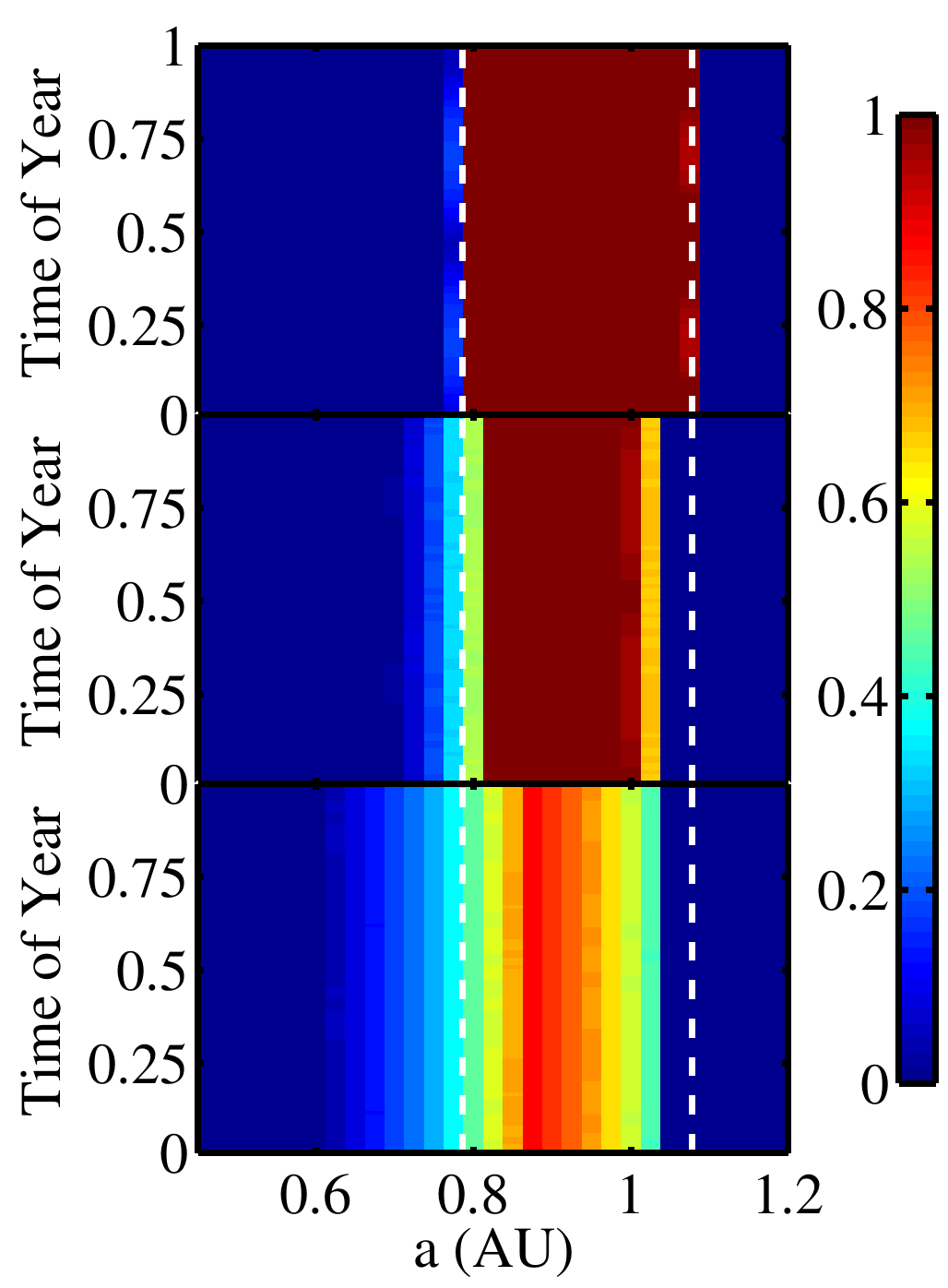}
\caption{Regional habitability fraction, $f_{\rm area}$, as function
of orbital distance and time of year.  Results are shown for the same
nine pseudo-Earth models as in Fig.~\ref{fig:temp}, with the same
panel structure.  Vertical dashed lines indicate the orbital extent of
the radiative equilibrium habitable zone.  In models with efficient
latitudinal heat transport (top row), the regional habitability
fraction tends to be either 0 or 1.  In low transport efficiency
models (bottom row), the regional habitability fraction takes on more
intermediate values.}
\label{fig:space hab}
\end{figure}

Figure~\ref{fig:space hab} shows temporal variation in fractional
habitability for the same nine pseudo-Earth EBMs as before.  In this
figure, contours indicate the value of the regional habitability,
$f_{\rm area}$, as a function of orbital separation and time of year.
As before, time of year is measured in fraction of a year from the
northern winter solstice.  There is little temporal variability in the
nine models shown.  This is a consequence of the North--South symmetry
and the large thermal inertia of these models, with $70$\% ocean
fraction.  A model with North--South asymmetry in the distribution of
land and sea, or even a symmetrical model with a much reduced ocean
fraction, would show more significant variations in the value of
$f_{\rm area}$ with time.

In models with efficient latitudinal heat transport (top panels),
$f_{\rm area}$ tends to be either 0 or 1, taking on intermediate
values only over a very small range of orbital separations.  This is a
consequence of the latitudinal isothermality of these models:
essentiall the entire planetary surface is either above 273~K or below
it.  By contrast, models with inefficient latitudinal heat transport
(bottom panels) are habitable over 100\% of the planetary surface area
only in a small portion of the regionally habitable zone.  In fact,
models shown in the lower middle and lower right panels are never
habitable over the entire surface area of the planet.

\subsubsection{Pseudo-Earth with a Small Ocean Fraction}
\label{sssec:less ocean}

The presence of abundant oceans on Earth plays a critical role in
regulating its climate, but there is no fundamental reason why a
terrestrial planet should have Earth--like ocean covering.  The heat
capacity $C$ in equation~(\ref{eq:diffu eq1}) has units of $\rm
[time]\times[energy\,flux]/[temperature]$.  Multiplying $C$ by a
characteristic temperature and dividing by a characteristic flux,
therefore, provides an estimate of the characteristic timescale for
thermal relaxation:
\begin{equation}
t_{\rm relax} \sim \frac{C \times \bar{T}}{(S/4)\times (1-\bar{A})}.
\label{eq:t_relax}
\end{equation}
For Earth--like conditions, using $T\sim 300$~K, $S\sim1.4\times
10^{6}{\rm~erg~cm^{-2}~s^{-1}}$, and $\bar{A} \sim 0.5$, we find
\begin{equation}
t^{\rm E}_{\rm relax} \sim  \frac{0.0017 {\rm~K}}{1 {\rm~erg~cm^{-2}}} C {\rm~s}.
\label{eq:t_relax earth}
\end{equation}
In our EBM, we assign to the atmosphere a heat capacity that accounts
for the type of (thermally coupled) surface that lies beneath it.
Since the atmospheric heat capacity over land in our EBM is $C_l =
5.25\times10^9 {\rm erg~cm^{-2}~K^{-1}}$, the corresponding relaxation
timescale is roughly $t^{\rm E,l}_{\rm relax}\sim 9\times10^6{\rm~s}$,
or about 0.3 years -- between 3 and 4 months.  On the other hand, the
atmospheric heat capacity over ocean is 40 times larger because the
atmosphere is able to tap into the larger thermal inertia of a
mixed--wind layer at the ocean's surface. This results in a relaxation
timescale $t^{\rm E,o}_{\rm relax}$ over ocean that is slightly longer
than 10 years in our models.  Consequently, in any EBM of a planet
that is more than one part in 40 (2.5\%) uniformly covered by ocean,
the thermal relaxation timescale is approximately given by the ocean
relaxation timescale times the ocean fraction.

The pseudo-Earth EBMs presented so far, all with a uniform 70\% ocean
fraction, have long thermal timescales of nearly a decade ($\sim 8$
years).  By contrast, a model planet with only a 10\% ocean fraction
would have a thermal timescale that is closer to a year.  In such a
model, one would expect significantly larger seasonal variations in
temperature over the course of a year.

\begin{figure}[h!]
\plotone {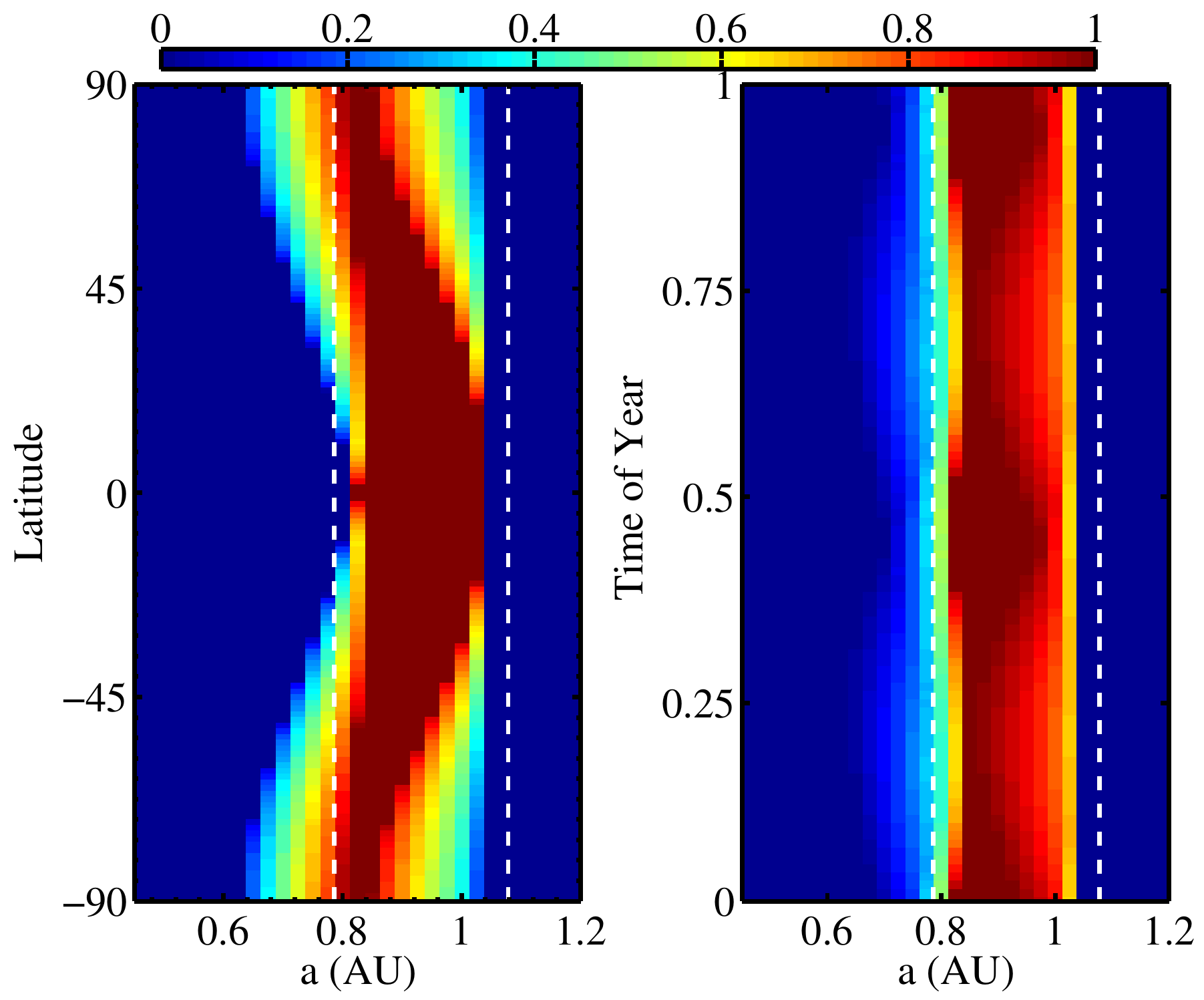}
\caption{Temporal and regional habitability fractions in a
pseudo-Earth model with a uniform 10\% ocean fraction.  The model is
in every way similar to our fiducial Earth-like model, except for the
reduced ocean fraction.  The left panel shows the temporal
habitability, $f_{\rm time}$, as a function of orbital distance and
latitude. The right panel shows the regional habitability, $f_{\rm
area}$, as a function of orbital distance and time of year. In each
panel, vertical dashed lines indicate the orbital extent of the
radiative equilibrium habitable zone.  With this smaller ocean
fraction, the temporal habitability (left panel) takes on more
intermediate values (between 0 and 1) and the regional habitability
(right panel) shows seasonal variations with time of year (two cycles
per year because of the North--South model symmetry).}
\label{fig:10p water}
\end{figure}

Figure~\ref{fig:10p water} shows temporal and regional habitability
diagrams for a pseudo-Earth EBM that is identical to our fiducial
Earth-like model, except for a uniform ocean fraction reduced from
70\% to 10\%.  The left panel shows the temporal habitability, $f_{\rm
time}$, as a function of orbital distance and latitude, while the
right panel shows the regional habitability, $f_{\rm area}$, as a
function of orbital distance and time of year (measured in fraction of
a year from the solstice).  In contrast to the fiducial model with
70\% ocean fraction shown in the central panels of
Figures~\ref{fig:time hab} and \ref{fig:space hab}, there are now
large swaths of parameter space, both in the left and right panels,
that take on habitability fractions between 0 and 1.  Seasonal
variations of the regional habitability fraction are easily identified
in the right panel, with two ``seasonal cycles'' per year from the
North--South symmetry of the model. Note that the range of orbital
distances with nonzero fractional habitability, for this drier
pseudo-Earth, differs only slightly from the corresponding range in
the fiducial Earth--like model with $70$\% ocean fraction, with an
outer edge somewhat extended.

Table~\ref{ta:HabZones} summarizes the different sizes of fractionally
habitable zones for the various models considered in this
study. Compared to global radiative balance calculations, fractionally
habitable zones extend closer to the star, simply as a result of the
range of regional conditions that obtain on the planet in our EBMs. On
the other hand, the outer reach of fractionally habitable zones tends
to be limited, before the outer radiative limit is reached, by
dynamical climate transitions to globally-frozen snowball states.
Planetary rotation rates and land-ocean fractions influence the
orbital extent of fractionally habitable zones.  Since these and other
relevant planetary attributes will be unknown for terrestrial
exoplanets, assessments of habitability for these newly discovered
worlds will be complicated by these uncertainties.

\section{On the Definition of Habitability}
\label{sec:def}

\renewcommand{\labelenumi}{(\alph{enumi})}
\renewcommand{\labelenumii}{(\roman{enumii})}

Determining the habitable zone of model or actual planets is an
interdisciplinary endeavor that requires the contributions of
biologists, climatologists, and astronomers.  In light of our results
on the rich phenomenology of habitable climates, we find it useful to
mention here a few qualitative points that may prove to be important
in refining the notion of habitability beyond the standard orbital
radius requirement for surface liquid--water generally discussed in
the astronomical literature (e.g., Kasting et al. 1993).\footnote{We
focus on water, rather than other potential solvents upon which a
biochemistry might be able to develop, because water is the only
molecule that we know can form the basis of a biosphere.}

Habitability on Earth is not strictly tied to the freezing point of
water.  \citet{carpenter_et_al2000} have discovered bacterial activity
in South Pole snow, indicating that some organisms can reproduce at
ambient temperatures below $-10\degr$~C and can live at temperatures
down to $-85\degr$~C.  Nor is habitability strictly tied to the
sea--level boiling point of water on Earth. \citet{kashefi+lovley2003}
have cultured a strain of archaea that happily reproduces at
$121\degr$~C, and remains viable at temperatures up to at least
$130\degr$~C.  The idea that life requires temperatures appropriate
for liquid water at 1~Atm pressure is therefore at least partially
flawed.  These examples would indicate that the habitable temperature
range ought to be extended to 263~K--394~K, or even possibly
188~K--403~K.  Biologists will have the final say on three vital
ranges of temperature, by determining when life can arise, survive and
reproduce.  The temperature regime in which organisms can reproduce
may be the relevant one for the definition of habitability in the
context of astronomical searches.

Even if life requires liquid water, one must still carefully consider
the upper temperature limit to habitability.  Water boils at 373~K
only for a surface pressure of 1~Atm.  There is so much water in the
Earth's oceans that the surface temperature could be well above 373~K
and the atmospheric H$_2$O pressure change could allow liquid water to
remain on the surface.  On the other hand, since water is a greenhouse
gas and humidity increases with temperature, a planet's ability to
cool is likely to become significantly reduced at some critical
temperature.  Unless additional feedbacks intervene at these high
temperatures, such as higher albedo from increased cloud coverage, a
regime of ``runaway greenhouse effect'' may be reached.
\citet{kasting1988} and \citet{sugiyama_et_al2005} have argued that
this runaway regime would be reached for global surface temperatures
between $\sim350$~K and $\sim400$~K on Earth.  These important
considerations, which have not been addressed in our modeling
strategy, suggest that the high--temperature end of the habitable
range may be the critical temperature for runaway greenhouse effect,
at least for Earth-like planets.

We note, however, that all these notions should also be fully
integrated with the regional and seasonal character of climate
systems.  Mean planetary conditions do not necessarily determine the
ability of a planet to host life. Rather, the various conditions
achieved on different parts of the planet, at different times of year,
do.  Knowning mean conditions is useful only insofar as they are a
good proxy for actual local conditions, but in many cases they may not
be.  Allowing for seasonality renders concepts such as the runaway
greenhouse effect significantly more complicated.  For example, if
parts of a planet (e.g., equatorial regions) are above the critical
temperature for runaway greenhouse effect, while other parts (e.g.,
polar regions) are not, determining whether or not the planet's
temperature does run away becomes a non-trivially coupled problem.

Finally, we note that in order for a planet to be considered
habitable, it may need attributes such as its net fractional
habitability, $f_{\rm hab}$, to exceed some minimum threshold
values. For one thing, this may be required for life to arise and
survive on the planet. But such a requirement may also be important in
terms of observational searches for biomarkers and other signatures of
life.  If only a small fraction of the planet's surface area is
habitable, or if it is habitable only for a small fraction of the
year, the planet may host life but this life may be unable to generate
large enough signatures -- e.g., in spectral biomarkers
\citep{kaltenegger+selsis2007,kiang_et_al2007,grenfell_et_al2007,seager_et_al2005}
-- to permit reliable detections with specific remote sensing
instruments. Understanding the minimum fractional habitability
necessary to produce sufficient atmospheric concentrations of
biomarkers detectable by instruments such as \textsl{TPF} and
\textsl{Darwin} will be an interesting challenge for the emerging
field of astrobiology.

\section{Conclusion}
\label{sec:conc}

We have reconsidered the notion of habitability for Earth-like planets
with seasonal energy-balance climate models. These models show that
the concept of regional and seasonal habitability is generally
important to assess the ability of terrestrial exoplanets to host
life. We find that previous evaluations of habitable zones may have
omitted important climatic conditions by focusing on close Earth
analogies. We illustrate this with two specific examples:
pseudo--Earths rotating at different rates or possessing a smaller
ocean fraction than Earth itself. These pseudo-Earths have
quantitatively different climatic habitability properties than the
Earth itself.

Comparisons to global radiative balance calculations show that
seasonal habitability generally extends the inner orbital range of
Earth-like planet habitable zones. The outer orbital range is reduced
relative to what global radiative balance calculations would indicate,
however, because the climate generally makes a dynamical transition to
a globally-frozen snowball state before the outer radiative limit is
reached. The stability of a planet's climate against snowball events
therefore has a strong impact on its habitability. Since this
stability is partly determined by external forcing factors such as the
magnitude of insolation and the length of winters, we expect issues of
climate dynamics to be central in determining the habitability of
terrestrial exoplanets, particularly if their forcing conditions are
generally different from the moderate cases encountered in the Solar
System.

\section*{Acknowledgments}
We acknowledge many helpful conversations with James Cho, Michael
Allison and Anthony Del Genio.  We thank Diana Spiegel for help
understanding the MIT GCM data.  We thank our referee, Manfred Cuntz,
whose careful reading of earlier drafts of the manuscript helped to
clarify the presentation in the final version.  CS acknowledges the
funding support of the Columbia Astrobiology Center through Columbia
University's Initiatives in Science and Engineering, and a NASA
Astrobiology: Exobiology and Evolutionary Biology; and Planetary
Protection Research grant, \# NNG05GO79G.

\appendix
\section*{APPENDIX}
\label{sec:appendix}
In this appendix, we describe several numerical tests developed in the
process of building our EBMs.  We discuss the importance of sufficient
spatial resolution for converged results and the effects that initial
conditions have in determining the final periodic climate solutions
obtained.

\subsection*{I Numerical Resolution}
\label{sec:res}

\begin{figure}[h!]
\plotone
{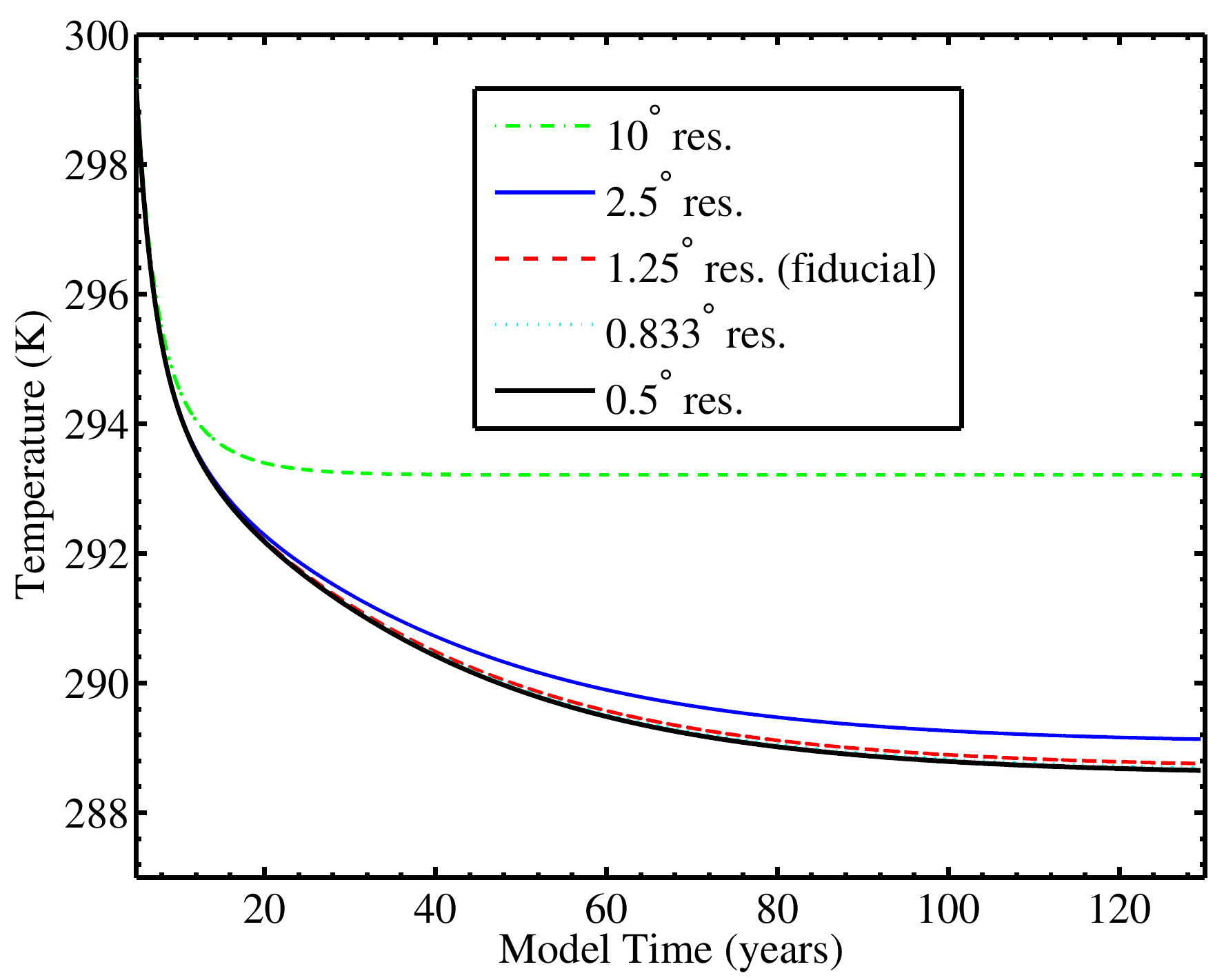}
\caption{Numerical convergence tests.  The global mean temperature in
our fiducial Earth-like model is shown, convolved with a 1--year
boxcar filter, for five different numerical resolutions. From top to
bottom, the resolution is $10\degr$ on the sphere (19 latitudinal grid
points), $2.5\degr$ (73 points), $1.25\degr$ (145 points),
$0.833\degr$ (217 points), and $0.5\degr$ (361 points). The default
resolution adopted in our work is $1.25\degr$.}
\label{fig:res comp}
\end{figure}

We investigated numerical resolution issues with a hot--start version
of our reference Earth-like model. It has uniform initial temperature
$T=350 {\rm~K}$, uses ``model~2'' formulations for the IR-cooling and
albedo functions (see Table~\ref{tab:one}) and adopts the fiducial
value $D=D_{\rm fid}$ for the magnitude of latitudinal heat
transport. Here, we describe the results of a convergence test for
five spatial resolutions: $10\degr$ on the sphere (19 grid points
equally spaced in latitude, including the poles); $2.5\degr$ (73
points); $1.25\degr$ (145 points); $0.833\degr$ (217 points); and
$0.5\degr$ (361 points; the ``high resolution model'').
Figure~\ref{fig:res comp} shows the time evolution of the global mean
surface temperature in the same Earth-like models at different
numerical resolutions.  Small-amplitude annual oscillations in the
global mean temperature have been smoothed out by convolving each
temperature time series with a 1--year boxcar filter.

As the numerical resolution is increased, differences between the
global temperature relaxation curves diminish rapidly, as expected
from the second-order spatial differencing scheme used. After 130
years, the model with $2.5\degr$ resolution achieves a periodic
solution for the global mean temperature that is within $\sim 0.5
{\rm~K}$ of the solution with the high resolution model. The
$1.25\degr$ resolution model stays within 0.2~K of the high resolution
model and the $0.833\degr$ resolution model is barely distinguishable
from the high resolution model (within $0.1$~K).  To strike a good
balance between numerical accuracy and integration speed, we adopt a
$1.25\degr$ resolution in all our models. This guarantees that our
results are not significantly affected by finite resolution
effects. With this level of numerical accuracy, it is clear that the
simplicity of various assumptions in our modeling strategy will be the
key factor limiting the predictive power of our EBMs.

We note that the model with $10\degr$ resolution relaxes to a periodic
solution much faster than the other models and that it settles to a
global mean temperature that deviates from the higher resolution
results by more than $4$~K. This consequence of poor numerical
resolution may be important since several EBM studies have been
carried out in the past at this standard $10\degr$ resolution (e.g.,
\citet{sellers1969}, WK97). Traditionally, one of the main
justifications for using such a low spatial resolution is that the
diffusive model for latitudinal heat transport is not expected to
provide a reliable description of much smaller spatial scales given an
Earth-like circulation regime (e.g.,
\citet{lorenz1979,north+coakley1979}).  It is our opinion, however,
that it is preferable to reach satisfactory numerical convergence with
the diffusive model even if the model itself is not be used to
interpret phenomena below some intermediate scale effectively
exceeding the numerical resolution. Otherwise, even global results
like the mean surface temperature shown in Fig.~\ref{fig:res comp} may
be prone to finite resolution effects. This becomes particularly
important in the context of climate dynamical studies where slight
differences in model attributes, such as initial conditions, can have
a profound influence on the ultimate outcome for the climate, as we
shall now illustrate.

\subsection*{II Initial Conditions}
\label{sec:init}

\begin{figure}[h!]
\plottwo
{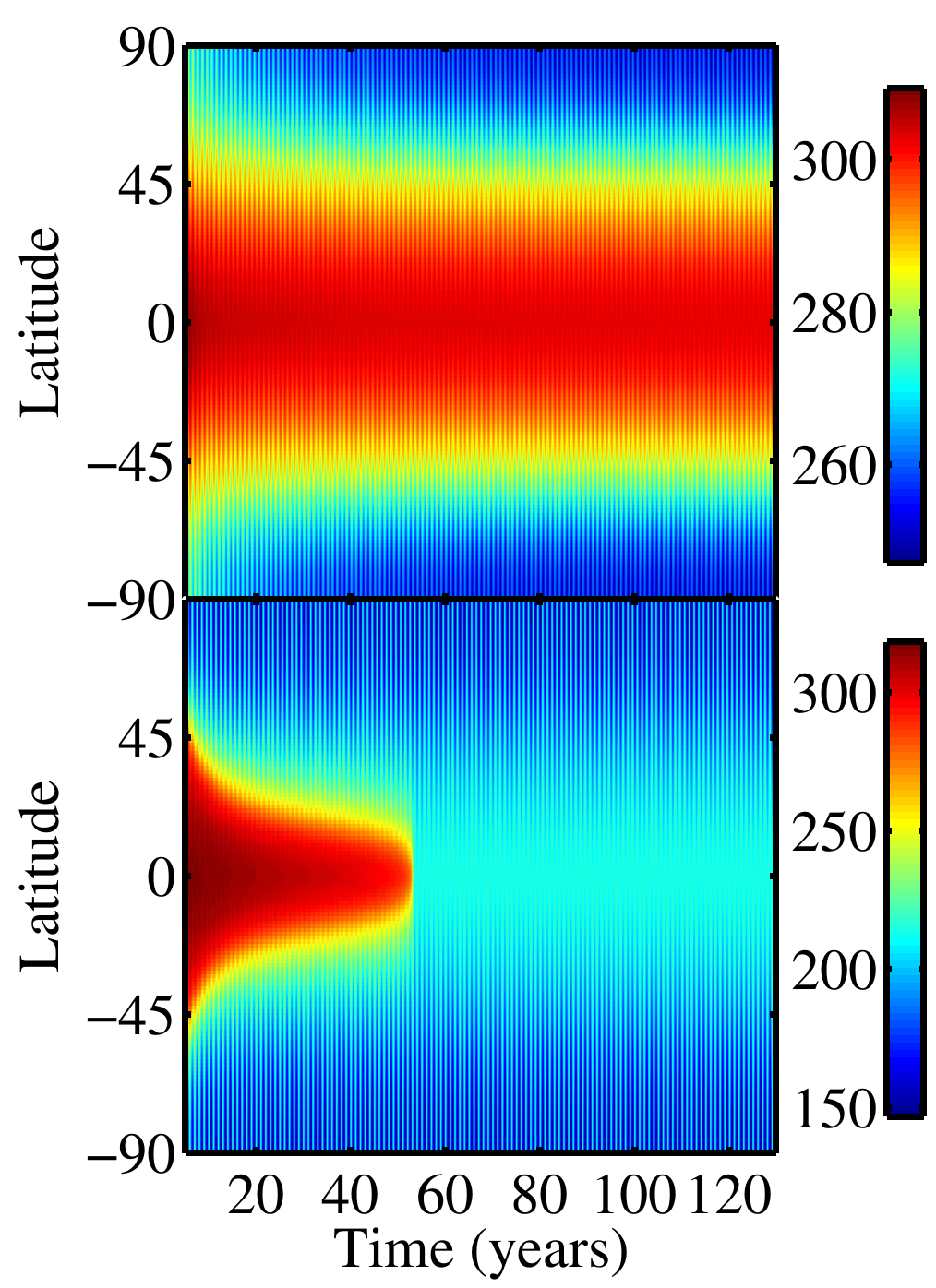} 
{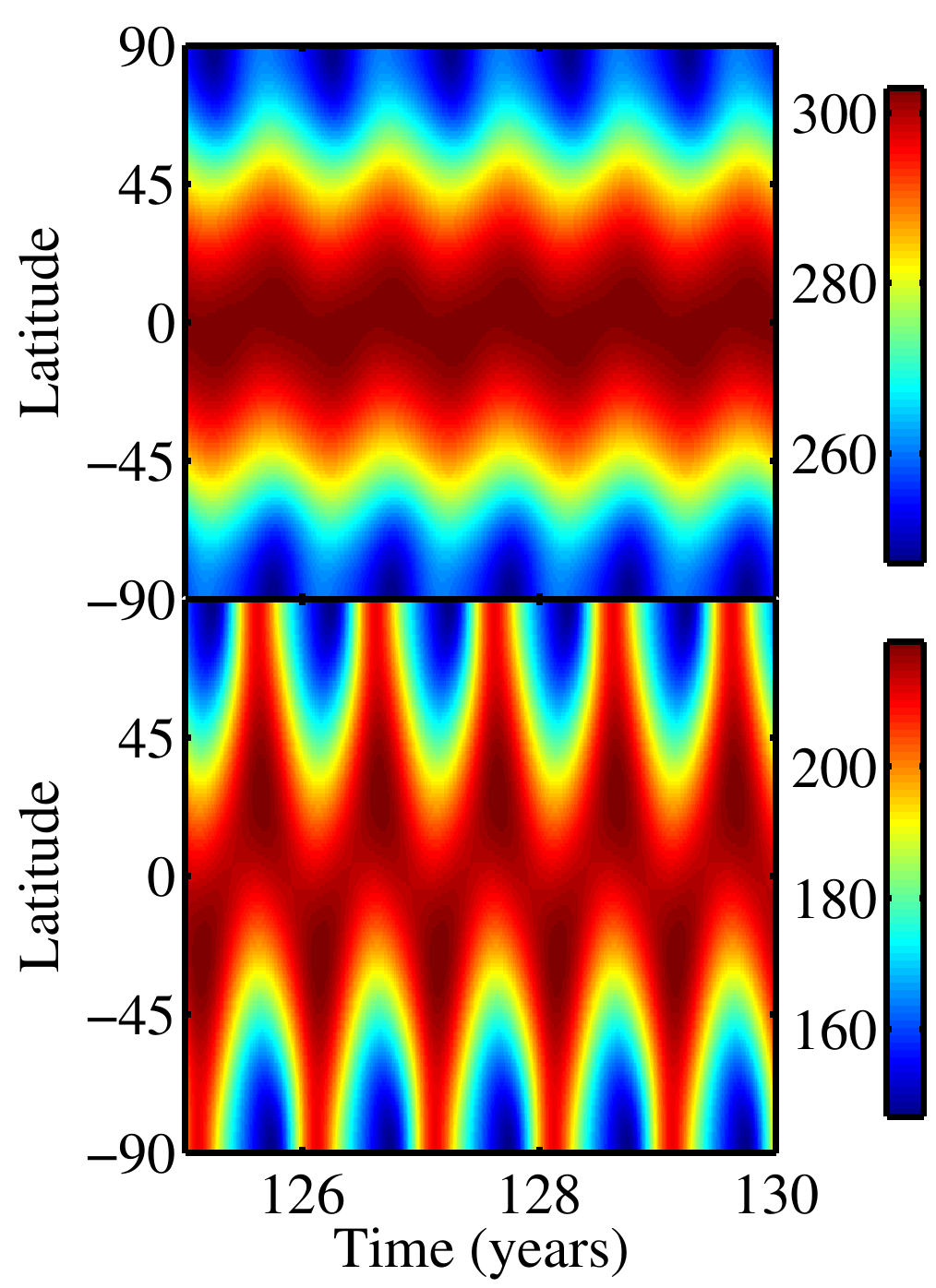} 
\caption{Space-time diagrams of temperature (in K) for two different
pseudo-Earth models.  Left panels show the entire 130 years of
evolution while the right panels show only the last 5 years of
seasonal climate. The top two panels show the evolution of
temperatures in our fiducial Earth-like model. The bottom two panels
show the evolution of temperatures in a pseudo-Earth model that is
identical to the fiducial model, except for a coefficient of
latitudinal heat transport reduced by a factor of 9 from the fiducial
value ($D_{\rm fid}/9$). A reduced efficiency of latitudinal heat
transport is expected on fast-rotating pseudo-Earths.  The model with
inefficient latitudinal heat transport makes a dynamical transition to
a globally-frozen snowball state after approximately $50$ years.}
\label{fig:temp maps}
\end{figure}

\begin{figure}[h!]
\plotone
{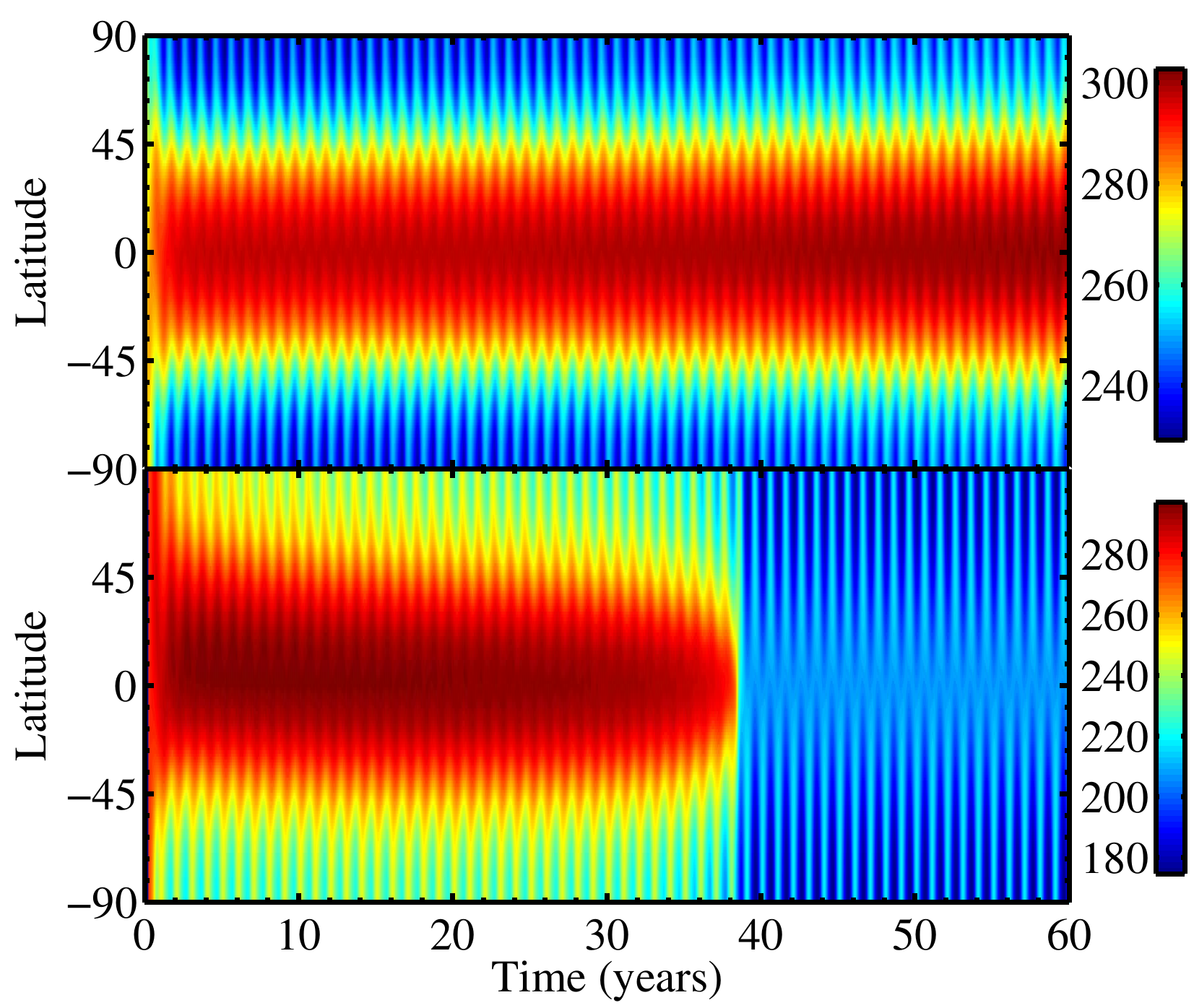}
\caption{Dependence on initial conditions (orbital time).  Both panels
show space-time diagrams of temperature in our fiducial Earth-like
model starting with a uniform temperature of $275$~K.  The model in
the top panel begins at the Northern winter solstice and it recovers
from the rather cool initial condition by asymptotically approaching
the same partially ice-covered climate solution as for typical
hot--start cases.  The model in the bottom panel, however, begins at
the Northern vernal equinox and makes a dynamical transition to a
globally-frozen snowball state after approximately 38 years.}
\label{fig:start time}
\end{figure}

Repeated experiments with our reference Earth-like EBM show that, as
long as the uniform initial temperature is chosen to be far above
$273$~K, essentially all memory of the initial conditions is lost by
130 years of evolution.  For initial temperatures approaching $273$~K,
however, there can be very sensitive dependence on the initial
conditions, due to strong water-ice albedo feedback effects.  Here we
show that even for the same initial conditions in temperature,
different outcomes for the climate are possible depending on the time
of year at which the model is started.  Figure~\ref{fig:start time}
shows space-time diagrams of temperature in two models which are
identical in all respects, including a uniform initial $T=275
{\rm~K}$, except for the initial time on the orbit.  The model in the
top panel begins with the $23.5$\degr\--obliquity planet at the northern
winter solstice while the one in the bottom panel begins at the
northern vernal equinox (a quarter year later).  After 60 years, the
model in the top panel is still evolving, but it is clearly making its
way to the same periodic, partially ice-covered solution as in the top
panels of Figure~\ref{fig:temp maps}. The model in the bottom panel,
on the other hand, transitions to a fully ice-covered snowball state after
slightly less than 40 years.  The North--South asymmetry evident in
the bottom panel shows that the model never lost memory of its
unbalanced start.  The model's northern hemisphere started in the
spring, and therefore experienced 6 months of greater insolation than
the southern hemisphere, from the very beginning.  As a result, the
southern hemisphere remained significantly, and increasingly, colder
than the northern hemisphere, until the cold temperatures in the south
were able to drag the entire planet down to a globally-frozen snowball
state.

\clearpage 

\begin{deluxetable}{cll}
\tablewidth{0pt}
\tablecaption{Atmospheric Models}
\tablehead{
\colhead{Model}  & \colhead{IR Cooling Function}  & \colhead{Albedo Function} 
}
\startdata
1\tablenotemark{a}  &   $I_1[T]  =  \frac{\sigma T^4}{1 + (3/4)\tau^0_{\rm IR}}$  &  $A_1[T]  =  0.5 - 0.2 \tanh[\frac{(T-268{\rm K})}{5{\rm K}}]$   \\
2\tablenotemark{b}  &  $I_2[T]  =  \frac{\sigma T^4}{1 + (3/4)\tau_{\rm IR}[T]}$   &  $A_2[T] =  0.525 - 0.245 \tanh[\frac{(T-268{\rm K})}{5{\rm K}}]$   \\
3\tablenotemark{c}  &   $I_3[T]  =  A + B T$   &  $A_3[T] =  0.475 - 0.225 \tanh[\frac{(T-268{\rm K})}{5{\rm K}}]$  \\
\enddata
\tablenotetext{a}{\,Model with fixed optical thickness: $\tau^0_{\rm IR} = 1$.}
\tablenotetext{b}{\,Model with $T$--dependent optical thickness: $\tau_{\rm IR}[T] = 0.79(T/273{\rm K})^3$.}
\tablenotetext{c}{\,Linearized model: $A = 2.033\times 10^5{\rm~erg~cm^{-2}~s^{-1}}$,
$B = 2.094\times 10^3 {\rm~erg~cm^{-2}~s^{-1}~K^{-1}}$.}
\tablecomments{$\sigma$ is the Stefan-Boltzmann constant.}
\label{tab:one}
\end{deluxetable}

\begin{deluxetable}{l||ccc}
\tablecaption{Habitable Zone Extents}
\tablehead{\colhead{Table~\ref{tab:one} Model:} & \colhead{\# 1} & \colhead{\# 2} & \colhead{\#3}}
\startdata
Global Radiative Balance      & 0.616-1.103 & 0.748-1.077 & 0.786-1.078 \\
EBM with $D=9D_{\rm fid}$     & 0.550-1.125 & 0.700-1.100 & 0.750-1.100 \\
EBM with $D=D_{\rm fid}$      & 0.450-1.075 & 0.625-1.025 & 0.675-1.050 \\
EBM with $D=(1/9)D_{\rm fid}$ & 0.400-1.100 & 0.525-1.000 & 0.575-1.050 \\
EBM with 10\% ocean fraction  & --          & 0.625-1.050 & -- \\
\enddata
\label{ta:HabZones}
\vspace{-0.4cm} \tablecomments{Habitable zones in each of the models
considered.  All orbital distances are in AU.}
\end{deluxetable}

\clearpage

\clearpage
\bibliography{biblio.bib}

\end{document}